\documentclass[12pt]{article}
\usepackage{amsmath,bbm,amssymb}
\usepackage{amsthm}
\usepackage{comment}
\usepackage{graphicx, graphics}
\usepackage{subcaption} 
\usepackage{textcomp}
\usepackage{amsfonts}
\usepackage{bm}
\usepackage{mathrsfs}
\usepackage{psfrag,pst-node,rotating, amsmath, bbm, amsthm, amssymb, amsthm, setspace, picture, epsfig, amsfonts, upgreek}
\usepackage{siunitx}
\usepackage{multirow}
\usepackage{graphicx}  
\usepackage{arydshln}
\usepackage{algorithm}
\usepackage{algorithmic}
\usepackage{subcaption}
\usepackage{arydshln}
\usepackage{enumitem}
\usepackage{booktabs}
\usepackage{multirow}
\usepackage{amsmath}
\usepackage{tabularx} 
\usepackage{natbib}
\usepackage{array}     
\usepackage{siunitx}   
\usepackage{caption}     
\usepackage{changepage}  
\usepackage{longtable}      
\usepackage{setspace}
\usepackage{hyperref} 
\usepackage[labelformat=simple]{subcaption}
\newtheorem{theorem}{Theorem}
\newtheorem{lemma}{Lemma}

\newcommand{\E}{\mathbb{E}}

\newcommand{\var}{\operatorname{Var}}

\newcommand{\dto}{\stackrel{d}{\longrightarrow}}

\def\v{\boldsymbol}

\begin{document}
\begin{center}
        {\Large\bf Nonparametric inference with massive data via grouped empirical likelihood}
\\[2mm] Yongda Wang$^{1,2}$, Shifeng Xiong$^{2,1}$\footnote{Corresponding author, Email: xiong@amss.ac.cn}
		{\footnotesize\\ 1. School of Mathematical Sciences, University of Chinese Academy of Sciences, Beijing, China
        \\[1mm] 2. State Key Laboratory of Mathematical Science, Academy of Mathematics and Systems Science, Chinese Academy of Sciences, Beijing, China
        }
\end{center}

\vspace{1cm} \noindent{\bf Abstract}

To address the computational issue in empirical likelihood methods with massive data,  this paper proposes a grouped empirical likelihood (GEL) method. It divides $N$ observations into $n$ groups, and assigns the same probability weight to all observations within the same group. GEL estimates the $n\ (\ll N)$ weights by maximizing the empirical likelihood ratio. The dimensionality of the optimization problem is thus reduced from $N$ to $n$, thereby lowering the computational complexity. We prove that GEL possesses the same first order asymptotic properties as the conventional empirical likelihood method under the estimating equation settings and the classical two-sample mean problem. A distributed GEL method is also proposed with several servers. Numerical simulations and real data analysis demonstrate that GEL can keep the same inferential accuracy as the conventional empirical likelihood method, and achieves substantial computational acceleration compared to the divide-and-conquer empirical likelihood method. We can analyze a billion data with GEL in tens of seconds on only one PC.

\vspace{1cm} \noindent{{\bf KEY WORDS:}} distributed computation; estimating equation; nonparametric likelihood; two-sample mean problem.

\section{Introduction}

Empirical Likelihood (EL) was proposed by Owen \cite{owen1988empirical} as an alternative to the bootstrap for constructing confidence regions in nonparametric problems. Building on Owen's foundational work, Qin and Lawless \cite{qin1994empirical} extended EL to general estimating equation frameworks. DiCiccio et al. \cite{diciccio1991empirical} established that empirical likelihood methods are subject to Bartlett correction for confidence interval construction. EL has found broad applicability in quantile inference \cite{chen1993smoothed, zhou2003adjusted}, ROC curve analysis \cite{claeskens2003empirical,qin2006empirical,yang2012smoothed,liu2012semi}, and Gini index estimation \cite{qin2010empirical,peng2011empirical}. More recently, Schennach \cite{schennach2007point1} pioneered the integration of exponential tilting techniques, Chen et al. \cite{chen2021sample} addressed the challenges of complex survey design with scrambled responses, and Thorne \cite{thorne2015empirical} as well as Xu and Chen \cite{xu2018empirical} demonstrated EL's utility in detecting differential gene expression. Research on two-sample problems using empirical likelihood has also revived. Qin \cite{qin1994semi} developed a semi-empirical likelihood approach for inferring the difference between two population means. Jing \cite{jing1995two} subsequently demonstrated that this two-sample empirical likelihood method for mean differences is Bartlett-correctable. Now, the EL method has become a primary statistical tool for nonparametric inference. This prominence stems from its methodological advantages over conventional approaches, including: minimal parametric assumptions for data distributions; the capacity to construct data-driven confidence regions; and natural incorporation of auxiliary information.

Advances in technology have driven exponential growth in data collection and dataset sizes. Although computing resources are also increasing rapidly, they pale in comparison to the astonishing surge in data volume. It is widely recognized that in large-scale learning, due to the immense computational demands, estimation and inference are the two primary challenges. The standard EL method assigns an individual probability weight to each data point, and estimates them via an optimization problem of maximizing the empirical likelihood ratio, which does not have a closed-form. As a result, EL can be computationally intensive, especially when applied to large datasets. While the empirical likelihood method can be computationally demanding, especially with large datasets, potentially limiting its practical application, its fundamental importance for massive data analysis remains undeniable. This is largely due to its strong theoretical ties with traditional statistical models and its distinct strengths in statistical inference. The challenges posed by massive datasets are not unique to empirical likelihood, other cornerstone nonparametric inference methods also face significant computational hurdles. For instance, scalable methods for the bootstrap, such as the Bag of Little Bootstraps \cite{kleiner2014scalable}, have been proposed. Consequently, overcoming these computational hurdles is crucial in the context of EL.

A variety of methods have been developed to address the challenges of big data, some of which also incorporate EL. For instance, subsampling-based methods \cite{kleiner2014scalable, ma2015statistical, wang2018optimal}, divide-and-conquer approaches \cite{zhang2013communication, chen2014split, lian2018divide}, and sequential updating algorithms \cite{schifano2016online} are among the most prominent. In particular, Ma et al. \cite{ma2022statistical}, Liu and Li \cite{liu2023distributed}, and Zhou et al. \cite{zhou2023distributed} have proposed combining the divide-and-conquer strategy with EL to enhance scalability and efficiency in large-scale inference.

Note that, for $N$ observations, there are $N$ probability weights needed to estimate via optimization techniques in the standard EL method. We observe that, in large-sample settings, these probability weights all tend to be close to each other. It seems that we do not need so many parameters to fit the underlying distribution. In this paper, we propose a grouped empirical likelihood (GEL) method, which divides the $N$ observations into $n$ groups, and assigns the same probability weight to all observations within the same group. These $n$ weights are also estimated by maximizing the empirical likelihood ratio. It can be seen that the dimensionality of the optimization problem from $N$ to $n$, thereby lowering the computational burden. We prove that the GEL method possesses the same first order asymptotic properties as EL under the estimating equation settings and the classical two-sample mean problem.
A distributed GEL method is also proposed when more than one servers are available.

Numerical simulations demonstrate that, with easy grouping strategies, GEL can keep the same inferential accuracy as EL, and simultaneously achieves substantial computational acceleration compared to both EL and divide-and-conquer EL. For a billion data, We can run GEL in tens of seconds on only one PC. The distributed GEL method can further accelerate the operation. The proposed methods are also applied to real datasets of anthropometric
records and A/B tests.

The rest of the paper is organized as follows. In Section \ref{Empirical likelihood}, we briefly review the empirical likelihood. In Section \ref{GEL}, we explain GEL method in details and establish its theoretical property. Section \ref{TGEL} extends GEL to the two-sample mean problem. Section \ref{DGEL} introduces the distributed GEL framework for distributed computing environments. In Section \ref{NS} and Section \ref{Real}, we assess the performance of proposed method via extensive simulation studies and real data analyses. Section \ref{Con} presents some concluding remarks and future directions. All proofs are deferred to the Appendix.

\section{Empirical likelihood} \label{Empirical likelihood}

In this section we briefly review the empirical likelihood (EL) method. Some notation and definitions are needed. For two vectors $\v{a} = (a_1, \dots, a_d)^{'}$ and $\v{b} = (b_1, \dots, b_d)^{'}$ in $\mathbb{R}^d$, the inequalities $\v{a} < \v{b}$ and $\v{a} \leq \v{b}$ are defined to hold if and only if $a_j < b_j$ and $a_j \leq b_j$ for all $j=1, \dots, d$, respectively. For a $d$-dimensional random vector $\v{X}$, let
$F(\bm{x}) = \Pr(\v{X} \leq \bm{x})$ denote its cumulative distribution function, where $\bm{x}\in\mathbb{R}^d$. We use $F(\bm{x}-)$ to denote $\Pr(\v{X} < \bm{x})$, and $\Pr(\v{X} = \bm{x}) = F(\bm{x}) - F(\bm{x}-)$. Let $\v{X}_1,\ldots,\v{X}_N$ be independent random observations of $\v{X}$. The empirical cumulative distribution function of $\v{X}_1, \dots, \v{X}_N$ is defined as
\[
F_N(\bm{x}) = \frac{1}{N}\sum_{i=1}^{N}\mathbb{I}\{\bm{X}_i \leq \bm{x} \}
\]
for $\bm{x}\in\mathbb{R}^d$, where $\mathbb{I}$ represents the indicator function. The nonparametric likelihood of any probability measure $F$ on $\mathbb{R}^d$ is defined as
\[
L(F)= \prod_{i=1}^{N}p_i = \prod_{i=1}^{N}(F(\bm{X}_i) - F(\bm{X}_i-)),
\]
where $p_i=F(\bm{X}_i) - F(\bm{X}_i-) \geq 0$ for $i = 1,\ldots,N$ and $\sum_{i=1}^{N}p_i = 1$.
Define $\mathcal F$ as the class of all $d$-dimensional cumulative distribution functions. It is known that
\begin{equation}\label{ecdf}\sup \{L(F): F \in \mathcal F\}=L(F_N) = N^{-N}.\end{equation}

The problem of interest is to make inference on a $p$-dimensional parameter $\bm{\theta}$ defined as the unique solution to the following estimating functions
\begin{equation} \label{0}
\mathbb{E}[\v{g}(\bm{X}, \bm{\theta})] = \v{0},
\end{equation}
where \(\v{g}(\bm{X}, \bm{\theta}) = (g_1(\bm{X}, \bm{\theta}), \ldots, g_r(\bm{X}, \bm{\theta}))^{'}\) and $r \geq p$. By \eqref{ecdf}, the EL ratio at \(\bm{\theta}\) is defined as
\[
R(\bm{\theta}) = \sup \left\{ \prod_{i=1}^N Np_i \mid p_i \geq 0,\  \sum_{i=1}^N p_i = 1,\  \sum_{i=1}^N p_i \v{g}(\bm{X}_i, \bm{\theta}) = \v{0} \right\}.
\]
Using the method of Lagrange multipliers, the above maximization problem yields
\begin{equation}\label{pi}
p_i =  \frac{1}{N\{1 + \boldsymbol{\lambda}^{'} \v{g}(\bm{X}_i, \bm{\theta})\}},
\end{equation}
where \(\boldsymbol{\lambda} \in \mathbb{R}^r\) is the solution to the following equations
\begin{equation}\label{ne}
\frac{1}{N} \sum_{i=1}^N \frac{\v{g}(\bm{X}_i, \bm{\theta})}{1 + \boldsymbol{\lambda}^{'} \v{g}(\bm{X}_i, \bm{\theta})} = \v{0}.
\end{equation}
By \eqref{pi}, the empirical log-likelihood ratio statistic for \(\bm{\theta}\) is given by
\begin{equation} \label{2}
-2\log R(\bm{\theta}) = 2 \sum_{i=1}^N \log \left\{ 1 + \boldsymbol{\lambda}^{'} \v{g}(\bm{X}_i, \bm{\theta}) \right\}.
\end{equation}
The maximum EL estimate of \(\bm{\theta}\), denoted by \(\hat{\bm{\theta}}_{EL}\), is defined as the minimizer of $-2\log R(\bm{\theta})$. Under regularity conditions, Qin and Lawless \cite{qin1994empirical} showed that as $N \rightarrow \infty$,
\[
\sqrt{N} \left( \hat{\bm{\theta}}_{EL} - \bm{\theta}_0 \right) \overset{d}{\longrightarrow} \mathcal{N}(\v{0}, \v{V}),
\]
where $\bm{\theta}_0$ denote the true value of $\bm{\theta}$ and

\begin{equation} \label{V_}
\bm{V} = \left\{ \E\left( \frac{\partial \v{g}(\bm{X}, \bm{\theta}_0)}{\partial \boldsymbol{\bm{\theta}}^{'}} \right)^{'} \left( \E\left( \v{g}(\bm{X}, \bm{\theta}_0) \v{g}(\bm{X}, \bm{\theta}_0)^{'} \right) \right)^{-1} \E\left( \frac{\partial \v{g}(\bm{X}, \bm{\theta}_0)}{\partial \boldsymbol{\theta}^{'}} \right) \right\}^{-1}.
\end{equation}
Moreover, the statistic for testing
\begin{equation}\label{H_0}
H_0: \bm{\theta} = \bm{\theta}_0 \quad\text{versus}\quad H_1: \bm{\theta} \neq \bm{\theta}_0
\end{equation}
is $-2\log R(\bm{\theta}_0)$. If $\mathbb{E}\left[\v{g}(\bm{X}, \bm{\theta}_0)\v{g}^{'}(\bm{X}, \bm{\theta}_0)\right]$ is positive definite and the rank of $\mathbb{E}[\partial \v{g}(\bm{X}, \bm{\theta}_0)/\partial \bm{\theta}]$ is $p$, then
\[
-2\log R(\bm{\theta}_0) \stackrel{d}{\longrightarrow} \chi^2_p \quad \text{as } N \rightarrow \infty.
\]

\section{Grouped empirical likelihood} \label{GEL}

When $N$ is sufficiently large, the classical EL method often incurs substantial computational costs due to the requirement of solving large-scale nonlinear equations in \eqref{ne}. To reduce the costs, we propose the grouped empirical likelihood (GEL) method that involves much fewer unknown parameters.

Under the same setting in the previous section, GEL randomly divides the $N$ sample points $\{\v{X}_1,\dots,\v{X}_N\}$ into $n$ disjointed groups ${\mathcal G}_1,\ldots, {\mathcal G}_n$. Let $ d_i $ represent the number of sample points in ${\mathcal G}_i$ for $i = 1,\dots,n$. We require that the sample sizes in each group should be as equal as possible, i.e., they satisfy
\[
\max_{i,k \leq n , i \neq k}\left| d_i - d_k \right| = 1 \qquad  \text{and} \qquad  N = \sum_{i=1}^{n}d_i.
\]

For simplicity, in the following we assume that the $N$ observations
\(\{\v{X}_1,\dots,\v{X}_N\}\)
can be evenly partitioned into \(n\) groups of size \(m\). Rewrite them as $\{\v{X}_{ij}\}_{i=1,\dots,n}^{j=1,\dots,m}$. The GEL method assigns the same probability measure \(q_i\) to all sample points within each group $\mathcal G_i,\ i=1, \ldots,n$. Hence, there are only $n$, instead of $N$, unknown parameters in the GEL method.

For the parameter $\bm{\theta}$ satisfying the estimating equations in (\ref{0}), the GEL framework maximizes $\prod_{i=1}^n q_i^m$ subject to restrictions
\begin{equation}\label{1}
q_i \geq 0, \quad \sum_{i=1}^{n} q_i = \frac{1}{m}, \quad \sum_{i=1}^{n} q_i \sum_{j=1}^{m} \v{g}(\bm{X}_{ij}, \bm{\theta}) = \v{0}.
\end{equation}Let
\[
\mathcal{H} = m \sum_{i=1}^{n} \log q_i + t \left( \frac{1}{m} - \sum_{i=1}^{n} q_i \right) + \bm{\lambda}^{'} \left( \sum_{i=1}^{n} q_i \sum_{j=1}^{m} \v{g}(\bm{X}_{ij}, \bm{\theta}) \right),
\]
where $t$ and $\bm{\lambda} = (\lambda_1, \cdots, \lambda_r)^{'}$ are Lagrange multipliers. Taking derivatives with respect to $q_i$, we have
\[
\frac{\partial \mathcal{H}}{\partial q_i} = \frac{m}{q_i} - t + \bm{\lambda}^{'} \sum_{j=1}^{m} \v{g}(\bm{X}_{ij}, \bm{\theta}) = 0,
\]
\[
\sum_{i=1}^{n}q_i\frac{\partial \mathcal{H}}{\partial q_i} = m + t \sum_{i=1}^n q_i + \bm{\lambda}^{'} \sum_{i=1}^n q_i \sum_{j=1}^m \v{g}(\bm{X}_{ij}, \bm{\theta}) = 0.
\]
We can get $t = -nm^2$ and
\begin{equation}\label{qi}
q_i = \frac{1}{N \left\{ 1 + \tilde{\bm{\lambda}}^{'} \bar{\v{g}}(\bm{X}_i, \bm{\theta}) \right\}},
\end{equation}
where $\bar{\v{g}}(\bm{X}_i, \bm{\theta}) = \sum_{j=1}^m \v{g}(\bm{X}_{ij}, \bm{\theta})/m$, $\tilde{\bm{\lambda}} = -\bm{\lambda}/N$, and for given $\bm{\theta}$, $\tilde{\bm{\lambda}}$ satisfies
\begin{equation}\label{lamda}
\frac{1}{n} \sum_{i=1}^n \frac{\bar{\v{g}}(\bm{X}_i, \bm{\theta})}{1 + \tilde{\bm{\lambda}}^{'} \bar{\v{g}}(\bm{X}_i, \bm{\theta})} = \v{0}.
\end{equation}

Since $0 \leq q_i \leq 1/m$, we have $1 + \tilde{\bm{\lambda}}^{'} \bar{\v{g}}(\bm{X}_i, \bm{\theta}) \geq 1/n$ for each $i$. Let $D_{\bm{\theta}} = \{ \tilde{\bm{\lambda}} : 1 + \tilde{\bm{\lambda}}^{'} \bar{\v{g}}(\bm{X}_i, \bm{\theta}) \geq 1/n \}$;, which is convex, closed, and bounded if $\v{0}$ is inside the convex hull of the $\bar{\v{g}}(\bm{X}_i, \bm{\theta})$ \cite{qin1994empirical}. Moreover,

\[
\frac{\partial}{\partial \tilde{\bm{\lambda}}} \left\{ \frac{1}{n} \sum_{i=1}^{n} \frac{\bar{\v{g}}(\bm{X}_i,\bm{\theta})}{1 + \tilde{\bm{\lambda}}^{'} \bar{\v{g}}(\bm{X}_i,\bm{\theta})} \right\} = - \frac{1}{n} \sum_{i=1}^{n} \frac{\bar{\v{g}}(\bm{X}_i,\bm{\theta}) \bar{\v{g}}(\bm{X}_i,\bm{\theta})^{'}}{\left\{ 1 + \tilde{\bm{\lambda}}^{'} \bar{\v{g}}(\bm{X}_i,\bm{\theta}) \right\}^2}
\]
is negative definite for $\tilde{\bm{\lambda}}\in D_{\bm{\theta}}$, provided that $\sum_{i=1}^{n} \bar{\v{g}}(\bm{X}_i,\bm{\theta}) \bar{\v{g}}^{'}(\bm{X}_i,\bm{\theta})$ is positive definite. By the inverse function theorem, $\tilde{\bm{\lambda}} = \tilde{\bm{\lambda}}(\bm{\theta})$ is thus a continuously differentiable function of $\bm{\theta}$.

The GEL function for $\bm{\theta}$ is now defined as
\[
L_G(\bm{\theta}) = \prod_{i=1}^{n} \left\{ \frac{1}{N \left\{ 1 + \tilde{\bm{\lambda}}^{'} \bar{\v{g}}(\bm{X}_i,\bm{\theta}) \right\}} \right\}^m.
\]
Since $\prod_{i=1}^{n} q_i^m$ is maximized at $q_i = N^{-1}$ without the constraints from the estimating equation \eqref{0}, the empirical log-likelihood ratio is
\begin{equation} \label{3}
-2\log R_G(\bm{\theta}) = 2m \sum_{i=1}^{n} \log \left\{ 1 + \tilde{\bm{\lambda}}^{'} \bar{\v{g}}(\bm{X}_i,\bm{\theta}) \right\}.
\end{equation}
We can obtain an estimate $\hat{\bm{\theta}}_{GEL}$ of $\bm{\theta}$ through minimizing $-2\log R_G(\bm{\theta})$. Consequently, $\tilde{\bm{\lambda}}$ can be estimated from (\ref{lamda}), and the estimate $\hat{q}_i$ of $q_i$ follows from (\ref{qi}), $i=1\ldots,n$. When $r = p$, $\hat{\bm{\theta}}_{GEL}$ is acturally the unconstrained maximum likelihood estimator, i.e., $\hat{q}_i = N^{-1}$. Algorithm \ref{alg:1} summarizes the procedure for computing $\hat{\bm{\theta}}_{GEL}$. Compared with the computation for $\hat{\bm{\theta}}_{EL}$ in the previous section, the proposed GEL method involves only $n\ll N$ parameters, and this significantly reduces the computational complexity.

\begin{algorithm}[t]
\caption{\quad GEL Estimation of $\bm{\theta}$ Algorithm.}
\begin{algorithmic}[1]\label{alg:1}
\STATE \textbf{Input:} Samples $\{\v{X}_{i}\}_{i=1}^{N}$; group numbers $n$; estimating functions $\v{g}(\v{X},\boldsymbol{\theta})$.

\STATE Randomly divide $\{\v{X}_{i}\}_{i=1}^{N}$ into $n$ groups $\{\v{X}_{ij}\}_{i=1,\dots,n}^{j=1,\dots,m}$ and compute $\bar{\v{g}}(\bm{X}_i, \bm{\theta}) = \sum_{j=1}^m \v{g}(\bm{X}_{ij}, \bm{\theta})/m$.

\STATE Obtain the GEL estimator via
\[
\hat{\bm{\theta}}_{GEL} = \mathop{\arg\min}\limits_{\theta \in \Theta}m \sum_{i=1}^{n} \log \left\{ 1 + \tilde{\bm{\lambda}}^{'} \bar{\v{g}}(\bm{X}_i,\bm{\theta}) \right\},
\]
where $\tilde{\bm{\lambda}}$ is the solution to the following equations:
\[
\frac{1}{n} \sum_{i=1}^n \frac{\bar{\v{g}}(\bm{X}_i, \bm{\theta})}{1 + \tilde{\bm{\lambda}}^{'} \bar{\v{g}}(\bm{X}_i, \bm{\theta})} = \v{0}.
\]

\STATE \textbf{Output:} $\hat{\bm{\theta}}_{GEL}$.

\end{algorithmic}
\end{algorithm}

We next prove that the GEL method possesses the same first order asymptotic properties as EL. Several conditions are needed. Let $\| \cdot \|$ denote the Euclidean norm.

\begin{enumerate}[label={}, leftmargin=*]
  \item \textbf{Condition 1.}
    $\mathbb{E}\left[\v{g}(\bm{X}, \bm{\theta}_0)\v{g}^{'}(\bm{X}, \bm{\theta}_0)\right]$ is positive definite.

  \item \textbf{Condition 2.}
   $\partial \v{g}(\bm{x}, \bm{\theta})/\partial \bm{\theta}$ and  $\partial^2\v{g}(\bm{x},\bm{\theta})/\partial\bm{\theta}\partial\bm{\theta}^{\top}$ are continuous in a neighborhood $\mathcal{N}(\bm{\theta}_0)$ of the true value $\bm{\theta}_0$.

  \item \textbf{Condition 3.}
   $\left\| \partial \v{g}(\bm{x}, \bm{\theta})/\partial \bm{\theta} \right\|$, $|| \partial^2\v{g}(\bm{x},\bm{\theta})/\partial\bm{\theta}\partial\bm{\theta}^{\top} ||$ and $\|\v{g}(\bm{x}, \bm{\theta})\|^3$ can be bounded by some integrable function $H(\bm{x})$ in the same neighborhood $\mathcal{N}(\bm{\theta}_0)$ as in Condition 2.

  \item \textbf{Condition 4.}
   The rank of $\mathbb{E}[\partial \v{g}(\bm{X}, \bm{\theta}_0)/\partial \bm{\theta}]$ is $p$.
\end{enumerate}

These conditions are generally mild and commonly used in the EL theory; see e.g., \cite{qin1994empirical, owen2001empirical}.

\begin{theorem} \label{thm:3.1}
Under Conditions 1-4, we have
\[
\sqrt{N} \left( \hat{\bm{\theta}}_{GEL} - \bm{\theta}_0 \right) \xrightarrow{d} \mathcal{N}\left( \v{0}, \bm{V}\right), \quad\text{as} \quad n \rightarrow \infty,
\]
where $\bm{V}$ is given by (\ref{V_}).
\end{theorem}

By (\ref{3}), the GEL ratio statistic for testing (\ref{H_0}) is $-2\log R_G(\bm{\theta}_0)$.
\begin{theorem} \label{thm:3.2}
Under Conditions 1-4 and $H_0$ in (\ref{H_0}), as $n \rightarrow \infty$, $-2\log R_G(\bm{\theta}_0)/m \xrightarrow{d} \chi_p^2$ when $H_0$ is true.
\end{theorem}

Proofs of Theorems \ref{thm:3.1} and \ref{thm:3.2} are given in the Appendix. It is worth noting that Theorems \ref{thm:3.1} and \ref{thm:3.2} do not require any condition on $m$. They hold for fixed $m$ or $m\to\infty$.
In addition, even with much fewer number of parameters, the convergence rate of the proposed estimators keeps the order of $1/\sqrt{N}$.

\section{Two-sample GEL}\label{TGEL}

Let $\v{X} \in \mathbb{R}^d$ and $\v{Y} \in \mathbb{R}^d$ be random vectors associated with parameters $\boldsymbol{\theta}_x \in \mathbb{R}^p$ and $\boldsymbol{\theta}_y \in \mathbb{R}^p$, respectively. Each parameter is defined as the unique solution to $p$ functionally independent unbiased estimating functions:
$$
\mathbb{E}\{\v{g}(\v{X}, \boldsymbol{\theta}_x)\} = \v{0}, \quad
\mathbb{E}\{\v{g}(\v{Y}, \boldsymbol{\theta}_y)\} = \v{0},
$$
where $\v{g}$ is a vector-valued function $\mathbb{R}^d\times\mathbb{R}^p\mapsto\mathbb{R}^p$.
Let $\boldsymbol{\pi} = \boldsymbol{\theta}_{y} - \boldsymbol{\theta}_{x}$ be the parameter of interest.

Suppose that $\{\v{X}_1,\dots,\v{X}_{N_1}\}$ is an i.i.d.\ sample from $\v{X}$ and
$\{\v{Y}_1,\dots,\v{Y}_{N_2}\}$ is an i.i.d.\ sample from $\v{Y}$, with the two samples mutually independent. Assume that both $N_1$ and $N_2$ are exact multiples of the block size $m$, i.e., $ N_1 = m\,n_1, N_2 = m\,n_2,$ for some integers $n_1,n_2\in\mathbb{N}$.  We then partition the two samples as $\{\v{X}_{ik}\}_{i=1,\dots,n_1}^{k=1,\dots,m}$ and $\{\v{Y}_{jt}\}_{j=1,\dots,n_2}^{t=1,\dots,m}$, respectively.

Define two probability vectors $\v{p}=(p_1,\dots,p_{n_1})^{'}$ and $\v{q}=(q_1,\dots,q_{n_2})^{'}$ satisfying $p_i\ge 0$, $q_j\ge 0$, and $\sum_{i=1}^{n_1}p_i=\sum_{j=1}^{n_2}q_j=1/m$.  For each $\v{p}$, let $\boldsymbol{\theta}_x(\v{p})$ solve
\begin{equation}\label{p}
\sum_{i=1}^{n_1} p_i \sum_{k=1}^{m} \v{g}(\v{X}_{ik},\boldsymbol{\theta}_x(\v{p})) = \v{0},
\end{equation}
and similarly define $\boldsymbol{\theta}_y(\v{q})$ by
\begin{equation}\label{q}
\sum_{j=1}^{n_2} q_j \sum_{t=1}^{m} \v{g}(\v{Y}_{jt},\boldsymbol{\theta}_y(\v{q})) = \v{0}.
\end{equation}
The two-sample GEL for $\boldsymbol{\pi}$ is defined as
\[
L_G(\boldsymbol{\pi}) = \sup
\Bigl(\prod_{i=1}^{n_1} p_i^m\Bigr)\Bigl(\prod_{j=1}^{n_2} q_j^m\Bigr),
\] subject to the constraints in (\ref{p}) and (\ref{q}).
The corresponding grouped empirical log-likelihood ratio is
\[
-2\log R_G(\boldsymbol{\pi}) = -2m
\Bigl\{  \sum_{i=1}^{n_1} \log(N_1 p_i) + \sum_{j=1}^{n_2} \log(N_2 q_j)\Bigr\}.
\]

Let $N = N_1 + N_2$, $\tau_1 = N/N_1$, and $\tau_2 = N/N_2$. According to \cite{liu2008empirical}, using the method of Lagrange multipliers, we have:
\[
-2\log R_G(\boldsymbol{\pi}) = 2m \left\{ \sum_{i=1}^{n_1} \log \left\{ 1 - \tau_1 ({\boldsymbol\lambda}^{*})^{'} \bar{\v{g}}(\boldsymbol{X}_i, \boldsymbol{\theta}_x^*)\right\} + \sum_{j=1}^{n_2} \log \left\{ 1 + \tau_2 ({\boldsymbol\lambda}^{*})^{'} \bar{\v{g}}(\boldsymbol{Y}_j, \boldsymbol{\theta}_y^*) \right\} \right\},
\]
where $\bar{\v{g}}(\boldsymbol{X}_i, \boldsymbol{\theta}_x^*) = \sum_{k=1}^{m} \v{g}(\boldsymbol{X}_{ik}, \boldsymbol{\theta}_x^*)/m$,
$\bar{\v{g}}(\boldsymbol{Y}_j,\boldsymbol{\theta}_y^*) = \sum_{t=1}^{m} \v{g}(\boldsymbol{Y}_{jt}, \boldsymbol{\theta}_y^*)/m$, and $(\boldsymbol{\theta}_x^*, \boldsymbol{\theta}_y^*,\boldsymbol{\lambda}^*)$ is the solution to the following nonlinear system:
\begin{equation}
\left\{
\begin{aligned} \label{6}
&\sum_{i=1}^{n_1} \frac{\bar{\v{g}}(\boldsymbol{X}_i, \boldsymbol{\theta}_x)}{1 - \tau_1 \boldsymbol{\lambda}^{'} \bar{\v{g}}(\boldsymbol{X}_i, \boldsymbol{\theta}_x)} = \v{0} \\
&\sum_{j=1}^{n_2} \frac{\bar{\v{g}}(\boldsymbol{Y}_j, \boldsymbol{\theta}_y)}{1 + \tau_2 \boldsymbol{\lambda}^{'} \bar{\v{g}}(\boldsymbol{Y}_j, \boldsymbol{\theta}_y)} = \v{0}. \\
&\boldsymbol{\pi} = \boldsymbol{\theta}_y - \boldsymbol{\theta}_x
\end{aligned}
\right.
\end{equation}Algorithm \ref{alg:2} outlines the steps for computing $-2\log R_G(\boldsymbol{\pi})$.

\begin{algorithm}[t]
\caption{\quad Algorithm for Calculating $R_G(\boldsymbol{\pi})$.}
\begin{algorithmic}[1]\label{alg:2}
\STATE \textbf{Input:}
  Samples $\{\v{X}_{ik}\}_{i=1,\dots,n_1}^{k=1,\dots,m}$,
          $\{\v{Y}_{jt}\}_{j=1,\dots,n_2}^{t=1,\dots,m}$;
  estimating functions $\v{g}(\v{X},\boldsymbol{\theta}_x)$, $\v{g}(\v{Y},\boldsymbol{\theta}_y)$.
\STATE Let $N_1=m n_1,\;N_2=m n_2,\;N=N_1+N_2,\;\tau_1=N/N_1,\;\tau_2=N/N_2$, $\bar{\v{g}}(\boldsymbol{X}_i, \boldsymbol{\theta}_x) = \sum_{k=1}^{m} \v{g}(\boldsymbol{X}_{ik}, \boldsymbol{\theta}_x)/m$, and
$\bar{\v{g}}(\boldsymbol{Y}_j, \boldsymbol{\theta}_y) = \sum_{t=1}^{m} \v{g}(\boldsymbol{Y}_{jt}, \boldsymbol{\theta}_y)/m$
\STATE Solve the nonlinear system of equations
\[
\sum_{i=1}^{n_1}\frac{\bar{\v{g}}(\v{X}_i,\boldsymbol{\theta}_x)}
     {1 - \tau_1\,\boldsymbol{\lambda}^{'}\bar{\v{g}}(\v{X}_i,\boldsymbol{\theta}_x)}
= \v{0},\quad
\sum_{j=1}^{n_2}\frac{\bar{\v{g}}(\v{Y}_j,\boldsymbol{\theta}_y)}
     {1 + \tau_2\,\boldsymbol{\lambda}^{'}\bar{\v{g}}(\v{Y}_j,\boldsymbol{\theta}_y)}
= \v{0},\quad
\boldsymbol{\theta}_y-\boldsymbol{\theta}_x=\boldsymbol{\pi}
\]
to get $(\boldsymbol{\lambda}^*,\boldsymbol{\theta}_x^*,\boldsymbol{\theta}_y^*)$.
\STATE \textbf{Output:} Compute
\[
-2\log R_G(\boldsymbol{\pi}) = 2m\Bigl\{
\sum_{i=1}^{n_1}\log\bigl(1 - \tau_1\,(\boldsymbol{\lambda}^*)^{'}\bar{\v{g}}(\v{X}_i,\boldsymbol{\theta}_x^*)\bigr)
+\sum_{j=1}^{n_2}\log\bigl(1 + \tau_2\,(\boldsymbol{\lambda}^*)^{'}\bar{\v{g}}(\v{Y}_j,\boldsymbol{\theta}_y^*)\bigr)
\Bigr\}.
\]
\end{algorithmic}
\end{algorithm}

For the hypotheses
\begin{equation}\label{4}
H_0: \boldsymbol{\pi} = \boldsymbol{\pi}_0 \quad\text{versus}\quad H_1: \boldsymbol{\pi} \neq \boldsymbol{\pi}_0,
\end{equation} the grouped empirical log-likelihood ratio statistic is $-2\log R_G(\boldsymbol{\pi}_0)$. To study its asymptotic properties, we need the following conditions.

\begin{enumerate}[label={}, leftmargin=*]
  \item \textbf{Condition 5.}
    For the true parameter values $\boldsymbol{\theta}_{x_0}$ and $\boldsymbol{\theta}_{y_0}$, the covariance matrices
    $\mathrm{Var}[\v{g}(\v{X},\boldsymbol{\theta}_{x_0})]$ and
    $\mathrm{Var}[\v{g}(\v{Y},\boldsymbol{\theta}_{y_0})]$
    are positive definite.

  \item \textbf{Condition 6.}
    \(\displaystyle \frac{\partial \v{g}(\bm{x},\boldsymbol{\theta}_x)}{\partial \boldsymbol{\theta}_x}\)
    and
    \(\displaystyle \frac{\partial^2 \v{g}(\bm{x},\boldsymbol{\theta}_x)}{\partial \boldsymbol{\theta}_x\,\partial \boldsymbol{\theta}_x^{'}}\)
    are continuous with respect to \(\boldsymbol{\theta}_x\) for
    \(\boldsymbol{\theta}_x\) in a neighborhood of \(\boldsymbol{\theta}_{x_0}\), and
    they are bounded by an integrable function of \(\bm{x}\).

  \item \textbf{Condition 7.}
    \(\displaystyle \frac{\partial \v{g}(\bm{y},\boldsymbol{\theta}_y)}{\partial \boldsymbol{\theta}_y}\)
    and
    \(\displaystyle \frac{\partial^2 \v{g}(\bm{y},\boldsymbol{\theta}_y)}{\partial \boldsymbol{\theta}_y\,\partial \boldsymbol{\theta}_y^{'}}\)
    are continuous with respect to \(\boldsymbol{\theta}_y\) for
    \(\boldsymbol{\theta}_y\) in a neighborhood of \(\boldsymbol{\theta}_{y_0}\), and
    they are bounded by an integrable function of \(\bm{y}\).

  \item \textbf{Condition 8.}
     \(\mathbb{E} \|\v{g}(\v{X}, \boldsymbol{\theta}_x)\|^{3} < \infty\) and \(\limsup_{\|t\| \to \infty} \left| \mathbb{E}[\exp\{i\v{t}^{'} \v{g}(\v{X}, \boldsymbol{\theta}_x)\}] \right| < 1\); \(\mathbb{E} \|\v{g}(\v{Y}, \boldsymbol{\theta}_y)\|^{3} \\ < \infty\) and \(\limsup_{\|t\| \to \infty} \left| \mathbb{E}[\exp\{i\v{t}^{'} \v{g}(\v{Y}, \boldsymbol{\theta}_y)\}] \right| < 1\), where $i$ denotes the imaginary unit.
\end{enumerate}

These conditions are commonly used in two-sample empirical likelihood studies \cite{tsao2015two}.

\begin{theorem}\label{thm:4.1}
Under Conditions 5-8 and $H_0$ in (\ref{4}), we have
\[
\frac{-2\log R_G(\boldsymbol{\pi}_0)}{m} \xrightarrow{d} \chi^2_p \quad \text{as} \quad n_1, n_2 \rightarrow \infty \text{ with } n_1/n_2 \rightarrow c \in (0,\infty).
\]
\end{theorem}

Theorem \ref{thm:4.1} holds with no restriction on $m$, whether it is fixed or tends to infinity. We defer the proof of Theorem \ref{thm:4.1} to the Appendix.

\section{Distributed GEL} \label{DGEL}

In the previous two sections, we have focused on the implementation of the GEL method on a single machine. However, in many practical scenarios involving massive datasets, data are often stored and processed in a distributed manner across multiple servers. In this section, we extend the GEL framework to such distributed settings.

Suppose the entire dataset of size $N$ is distributed across $K$ servers. Let $N_k$ denote the sample size on the $k$-th server with $N = \sum_{k=1}^K N_k$. For each $k$, the GEL method partitions the data into $n_k$ groups of size $m_k$, with equal weights assigned to observations within each group. Then, we apply the GEL method independently to the local data, obtaining a local parameter estimate $\hat{\boldsymbol{\theta}}_k$ and a local log-likelihood ratio statistic $-2\log R_{G,k}(\boldsymbol{\theta})$.
The distributed GEL (DGEL) estimator $\hat{\boldsymbol{\theta}}_{DGEL}$ is then computed by averaging the local estimates,
\[
\hat{\boldsymbol{\theta}}_{DGEL} = \frac{1}{K} \sum_{k=1}^K \hat{\boldsymbol{\theta}}_k.
\]
Similarly, for testing (\ref{H_0}), the overall log-likelihood ratio statistic can be aggregated as
\[
-2\log R_{DGEL}(\boldsymbol{\theta}) := \frac{1}{K}\sum_{k=1}^K -2\log R_{G,k}(\boldsymbol{\theta}).
\]

Under regularity conditions similar to those in Section 3, it can be shown that $\hat{\boldsymbol{\theta}}_{DGEL}$ is consistent and asymptotically normal, and $-2\log R_{DGEL}(\boldsymbol{\theta}_0)$ converges in distribution to a chi-square distribution. The DGEL strategy can also be used for the two-sample problem in Section \ref{TGEL}.

\section{Numerical simulation} \label{NS}

In this section we conduct numerical experiments to compare the performance of the following EL methods: the standard EL method \cite{qin1994empirical}, the DCEL method \cite{liu2023distributed}, and the proposed GEL and DGEL methods. The \texttt{emplike} function in the \texttt{statsmodels} package of \texttt{Python} is used to compute the EL and GEL estimators.

\textbf{Example 1} (Estimating the parameters of a normal distribution \cite{liu2023distributed}). In this example \( X_1, \ldots, X_N \) are independently drawn from the normal distribution \( \mathcal{N}(\mu, \sigma^2) \), where  \( \mu = 0 \) and \( \sigma = 2 \). Note that \( \mu \) and \( \sigma^2 \) satisfy the following moment conditions:
\[
\E \left[ \v{g} \left( X_1, \theta \right) \right] \equiv \E
\begin{bmatrix}
\mu - X_1 \\
\sigma^2 - \left( X_1 - \mu \right)^2 \\
X_1^3 - \mu \left( \mu^2 + 3\sigma^2 \right)
\end{bmatrix} = 0.
\]

We first compare the EL methods with $N = 10^5$ on a single server. In DCEL, we consider $k = 10, 100, 1000$, and $5000$, and divide the entire sample set into $k$ disjoint blocks $S_1,\ldots,S_k$ of (approximately) equal sizes. For each $S_j$, we compute the EL estimators $\hat{\mu}_j$ and $\hat{\sigma}_j$ based on the data in $S_j$. The final aggregated estimators are given by $\hat{\mu}_{DCEL} = \sum_{j=1}^k \hat{\mu}_j/k$ and $\hat{\sigma}_{DCEL} = \sum_{j=1}^k \hat{\sigma}_j/k$. In GEL, we set $m = 10, 100, 1000,$ and $5000$, which correspond to the effective sample size $n=10000, 1000, 100, 20$, and $10$, respectively.

Table \ref{tab:example1-results} presents the mean squared error (MSE) of each estimator based on 1000 replications, along with the average computational time for a single experiment across all three methods. It can be seen that, the effectiveness of DCEL in estimating $\sigma$ diminishes when $k$ equals 1000 and 5000, performing less favorably than both EL and GEL. With appropriate $m$, the proposed GEL method demonstrates satisfactory performance with comparable estimation accuracy to EL and DCEL but much higher computational speed. Furthermore, within the GEL framework, $n = N/m$ can be viewed as the effective sample size of the EL problem. Although a smaller value of $n$ leads to greater computational efficiency, $n$ must be sufficiently large to ensure the reliability of statistical inference. In this example $n\in[20,\ 100]$ seems to be a good balance.

\begin{table}[t]
\centering
\caption{MSE and average computation time in Example 1 (standard deviations in parentheses).}
\label{tab:example1-results}

\setlength{\tabcolsep}{8pt}
\renewcommand{\arraystretch}{1.2}

\begin{tabular*}{\textwidth}{@{\extracolsep{\fill}} l c c c  @{}}
\toprule
Method & $\mu\ (\times 10^{-5})$ & $\sigma\ (\times 10^{-4})$ & {Time (sec.)} \\
\midrule
EL   & 7.9615 (6.8961) & 6.8807 (5.9562)& 1.8712 \\
\midrule
DCEL $(k=10)$    & 7.9623 (6.8972)& 6.7976 (5.8714)& 0.1551 \\
GEL $(m=10, n=10000)$    & 7.9623 (6.8973)& 6.8824 (5.9623)& 0.0982 \\
\midrule
DCEL $(k=100)$    & 7.9623 (6.8972)& 6.6054 (5.3852)& 0.2717 \\
GEL $(m=100, n=1000)$    & 7.9623 (6.8972)& 6.8819 (5.9620)& 0.0639 \\
\midrule
DCEL $(k=1000)$   & 7.9684 (6.9423)& 8.5057 (8.4236)& 2.2119 \\
GEL $(m=1000, n=100)$   & 7.9623 (6.8972)& 6.8824 (5.9623)& 0.0038 \\
\midrule
DCEL $(k=5000)$  & 7.9683 (6.9420)& 14.2426 (12.3267)& 10.4574 \\
GEL $(m=5000, n=20)$  & 7.9615 (6.8892)& 6.8817 (5.9615)& 0.0028 \\
\midrule
DCEL $(k=10000)$  & 8.7986 (7.9656)& 16.5652 (14.9234)& 23.2565 \\
GEL $(m=10000, n=10)$  & 8.2615 (7.4654)& 8.3512 (7.2352)& 0.0017 \\
\bottomrule
\end{tabular*}
\end{table}

We next compare the proposed methods and DCEL under distributed settings with sample size $N \in \{10^6, 5\times10^6, 10^7, 5\times10^7, 10^8\}$. Here DCEL and DGEL are conducted in parallel across ten servers. GEL on only one server is also compared. For DCEL, we consider two settings with $k=1$ and $k=100$ on each server. For the proposed methods, $m$ is chosen such that the effective sample size $n=100$.

\begin{figure}[htbp]
\centering
\includegraphics[width=0.8\textwidth]{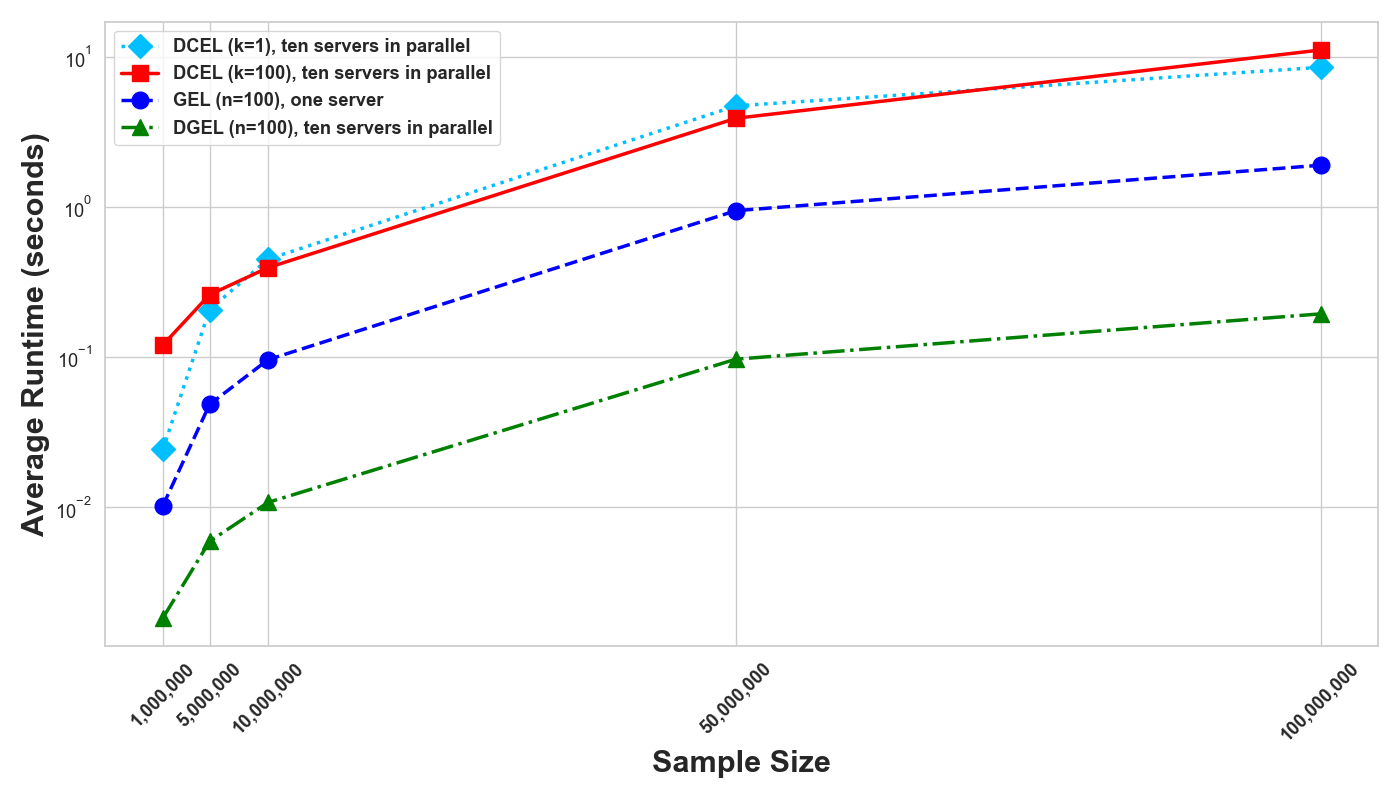}
\caption{Average computation time for GEL (single server), DCEL (ten servers), and Distributed GEL (ten servers) across varying sample sizes.}
\label{fig:runtime_comparison}
\end{figure}

Figure \ref{fig:runtime_comparison} presents the results of average computational time, demonstrating the significant advantages of the proposed GEL methods. Notably, for large sample sizes, even the single-server implementation of GEL substantially outperforms the 10-server parallel implementation of DCEL in terms of computational speed. Furthermore, when deployed across multiple servers, our DGEL method achieves the fastest performance, showcasing its superior scalability and efficiency for massive data analysis.

\vspace{\baselineskip}

\noindent\textbf{Example 2} (Linear regression with heteroscedastic noise). As demonstrated in Example 1, the speed advantage of our method is more obvious in a parallel computing environment with multiple servers. The subsequent experiments will be conducted on a single server. In this example we consider estimating the coefficients of the linear regression model:
\[
Y_i = \beta_0+\bm{X}_i^{'} \bm{\beta} + \epsilon_i, \quad i = 1, 2, \ldots, N,
\]
where $\bm{X}_i = (X_{i1}, \ldots, X_{ip})^{'}$ with $p \geq 5$, $\bm{\beta} = (\beta_1, \ldots, \beta_p)^{'}$ with the linear constraint $\sum_{j=1}^{5} \beta_j = 15$, and $\epsilon_i$ denotes the random error.  The covariates $\bm{X}_1,\ldots,\bm{X}_N$ are i.i.d. from $\mathcal{N}(\bm{0}, (1 - \rho)\bm{I}_p + \rho \v{1}_p)$, where $\rho\in(-1,1)$, $\bm{I}_p$ denotes the $p \times p$ identity matrix, and $\v{1}_p$ represents the $p$-dimensional column vector of ones. Given the covariates, the errors are generated from normal distributions with mean zero and
\[
\text{Var}(\epsilon_i \mid \bm{X}_i) = 1 + \alpha \left( \frac{\v{1}_p'\bm{X}_i}{\sqrt{p}} \right)^2, \quad i=1,\ldots,N,
\]
where $\alpha > 0$ controls the heteroscedasticity intensity. In the simulation we fix $N=10^5$, $\beta_0=1$, and $\bm{\beta} = (1, 2, \ldots, p)^{'}$, and consider three combinations of $(p,\rho,\alpha)$: $(5,0.2,1)$, $(10,0,2)$, and $(50,0.5,5)$.
Note that parameter vector \( \bm{\beta} \)  satisfies the following moment conditions:
\[
\E \left[ \v{g} \left( Y_1, \bm{X}_1, \bm{\beta} \right) \right] \equiv \E
\begin{bmatrix}
\bm{X}_1 (Y_1 - \bm{X}_1^{'} \bm{\beta}) \\
\sum_{j=1}^{5} \beta_j - 15
\end{bmatrix} = 0.
\]

\begin{table}[htbp]
\centering
\caption{MSE and average computation time in Example 2 (standard deviations in parentheses).}
\label{tab:example2-results}
\setlength{\tabcolsep}{24pt}
\renewcommand{\arraystretch}{1.2}
\footnotesize
\text{$\ p=5,\ \rho=0.2,\ \alpha=1$}\\[1em]
\begin{tabular}{l c c}
\toprule
Method &  MSE $(\times10^{-5})$ & Time (sec.) \\
\midrule
EL $(k=m=1)$ & 2.9095 (2.1992) & 0.8085  \\
\midrule
DCEL $(k=10)$ & 2.9362 (2.2089) & 0.3039  \\
GEL $(m=10, n=10000)$ & 2.8984 (2.1737) & 0.4024  \\
\midrule
DCEL $(k=100)$& 2.9375 (2.2274) & 0.1779  \\
GEL $(m=100, n=1000)$& 2.9087 (2.2135) & 0.0986  \\
\midrule
DCEL $(k=200)$& 2.9241 (2.2023) & 0.2503  \\
GEL $(m=200, n=500)$& 2.8963 (2.1884) & 0.0647  \\
\midrule
DCEL $(k=500)$& 2.9162 (2.2155) & 0.3015  \\
GEL $(m=500, n=200)$& 2.9038 (2.2142) & 0.0305  \\
\midrule
DCEL $(k=1000)$& 2.9855 (2.2174) & 0.4761 \\
GEL $(m=1000, n=100)$& 2.8772 (2.2005) & 0.0131  \\
\midrule
DCEL $(k=10000)$& 2.9379 (2.3201) & 674.85  \\
GEL $(m=10000, n=10)$& 3.8721 (6.2057) & 0.0060  \\
\bottomrule
\end{tabular}
\end{table}

\begin{table}[htbp]
\centering
\setlength{\tabcolsep}{24pt}
\renewcommand{\arraystretch}{1.2}
\footnotesize

\text{$p=10,\ \rho=0,\ \alpha=2$}\\[1em]
\begin{tabular}{l c c}
\toprule
Method &  MSE $(\times10^{-5})$ & Time (sec.) \\
\midrule
EL $(k=m=1)$ &  3.4055 (3.2043) & 1.3052  \\
\midrule
DCEL $(k=10)$&  3.3984 (3.1205) & 0.6599 \\
GEL $(m=10, n=10000)$&  3.4655 (3.2624) & 0.6457 \\
\midrule
DCEL $(k=100)$&  3.4715 (3.2329) & 0.2971\\
GEL $(m=100, n=1000)$&  3.3056 (3.2351)& 0.2056 \\
\midrule
DCEL $(k=200)$&  3.5034 (3.3275)& 0.3408 \\
GEL $(m=200, n=500)$&  3.2863 (3.2359)& 0.0853 \\
\midrule
DCEL $(k=500)$&  3.5047 (3.3245)& 0.6441 \\
GEL $(m=500, n=200)$&  3.3204 (3.2145)& 0.0265 \\
\midrule
DCEL $(k=1000)$&  3.5591 (3.3651)& 0.9276 \\
GEL $(m=1000, n=100)$&  3.4125 (3.3213)& 0.0047 \\
\midrule
DCEL $(k=10000)$& 26.984 (22.656)& 692.43\\
GEL $(m=10000, n=10)$& 14.865 (12.965)& 0.0035 \\
\bottomrule
\end{tabular}
\end{table}

\begin{table}[htbp]
\centering
\setlength{\tabcolsep}{24pt}
\renewcommand{\arraystretch}{1.2}
\footnotesize
\text{$p=50,\ \rho=0.5,\ \alpha=5$}\\[1em]
\begin{tabular}{l c c}
\toprule
Method &  MSE $(\times10^{-3})$ & Time (sec.) \\
\midrule
EL $(k=m=1)$& 2.4264 (2.3683) & 3.6597 \\
\midrule
DCEL $(k=10)$&  2.4251 (2.3694) & 3.0654 \\
GEL $(m=10, n=10000)$&  2.4249 (2.3687)& 2.6836 \\
\midrule
DCEL $(k=100)$&  2.6035 (2.4698)& 2.3278 \\
GEL $(m=100, n=1000)$&  2.4262 (2.3677)& 0.3559 \\
\midrule
DCEL $(k=200)$&  2.7132 (2.6987)& 2.9674 \\
GEL $(m=200, n=500)$&  2.4365 (2.3965)& 0.0853 \\
\midrule
DCEL $(k=500)$ &  3.2762 (2.9873)& 4.2926 \\
GEL $(m=500, n=200)$ &  2.4236 (2.3871)& 0.0922 \\
\midrule
DCEL $(k=1000)$ &  4.6365 (4.2745)& 4.4244 \\
GEL $(m=1000, n=100)$ &  2.4259 (2.3689)& 0.0638 \\
\midrule
DCEL $(k=10000)$ &  8.5032 (6.6541)& 4657.6 \\
GEL $(m=10000, n=10)$ & 6.9961 (5.9654)& 0.0314 \\
\bottomrule
\end{tabular}
\end{table}

The same three methods are compared as in Table \ref{tab:example1-results}. Table \ref{tab:example2-results} presents simulation results based on 1000 replications.
Like the findings in Example 1, with appropriate $m$, the proposed GEL method possesses comparable estimation accuracy to EL and DCEL but much shorter computational time. We also recommend selecting $m$ such that $n=N/m$ is of the order $10^2$. In addition, the performance of GEL is much more stable than DCEL in extreme cases of higher dimensions and/or strong heteroscedasticity.

\vspace{\baselineskip}

\noindent\textbf{Example 3} (Two-sample mean test) We consider a simulation study to evaluate the performance of the two-sample GEL method for testing the equality of population means.
The two populations are given by
\begin{align*}
X &\sim \frac{1}{3}N(0, 1) + \frac{1}{3}N(10^2, 10^2) + \frac{1}{3}N(10^3, 10^3), \\
Y &\sim \frac{1}{3}N(0, 2) + \frac{1}{3}N(10^2, 2 \times 10^2) + \frac{1}{3}N(10^3 + 20j, 3 \times 10^3).
\end{align*}
Here, $j$ is an integer index ranging from $0$ to $5$, which systematically introduces a shift in the mean of the third component in the distribution of $Y$.
We have $\mu_X=E(X) = 1100/3$ and $\mu_Y=E(Y) = (1100 + 20j)/3$, and the difference in means, $\Delta\mu = E(Y) - E(X)$, takes the values $\{0, 20/3, 40/3, 20, 80/3, 100/3\}$ for $j = 0, 1, \dots, 5$.
Consider the null hypothesis $H_0: \mu_X = \mu_Y$ against the alternative $H_1: \mu_X \neq \mu_Y$. Clearly, the case for $j=0$ corresponds to the null hypothesis.

For each $j$, we generate an i.i.d. sample of size $N_1 = N_2 = 30,000$ from each population.
The simulation is repeated for $1000$ times to assess the empirical Type I error rate for $j=0$ and statistical power for $j > 0$.
The significance level is set to $\alpha = 0.05$. For the two-sample GEL method, we evaluate its performance under different group sizes, setting $m_1 = m_2 = 100, 200, 300$, and $500$, which correspond to $n_1=n_2=300, 150, 100$, and $60$, respectively.

\begin{figure}[ht!]
  \centering
  \captionsetup{aboveskip=4pt, belowskip=0pt}
  \begin{subfigure}[b]{0.45\linewidth}
    \centering
    \includegraphics[width=\linewidth]{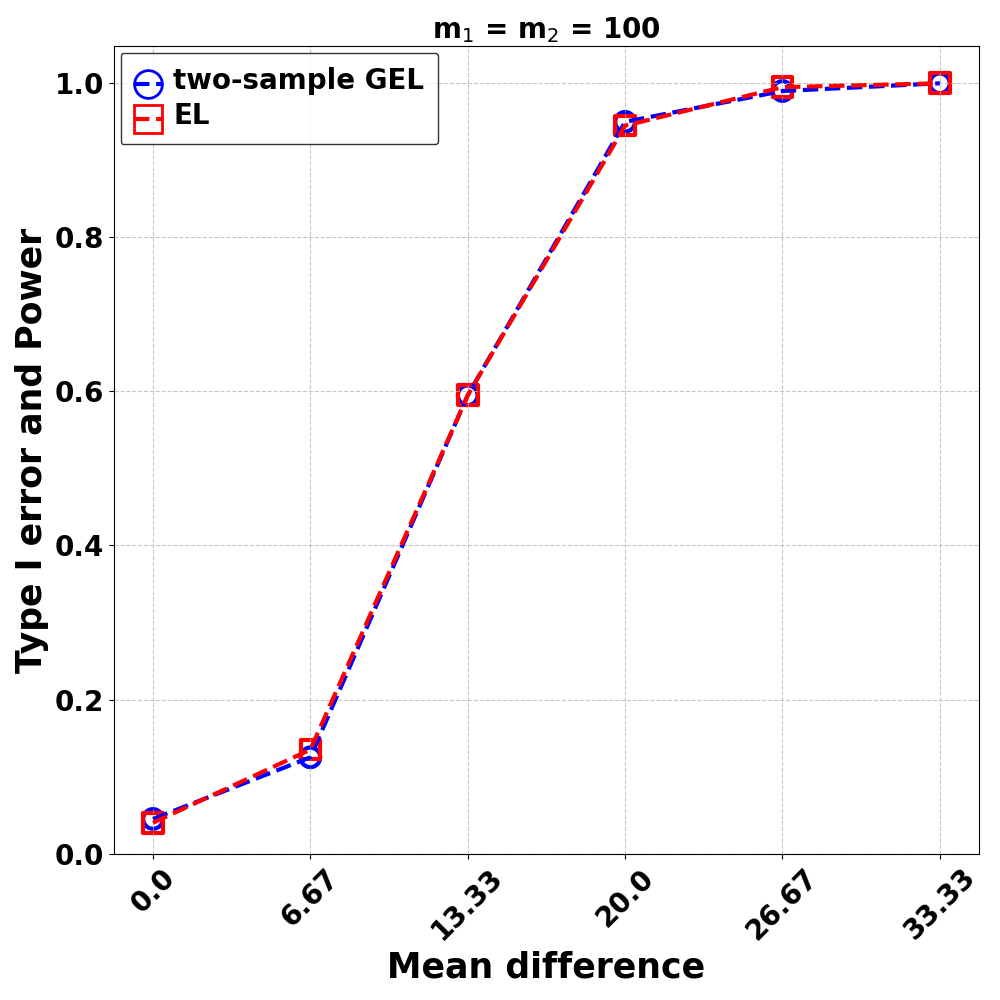}
  \end{subfigure}\hfill
  \begin{subfigure}[b]{0.45\linewidth}
    \centering
    \includegraphics[width=\linewidth]{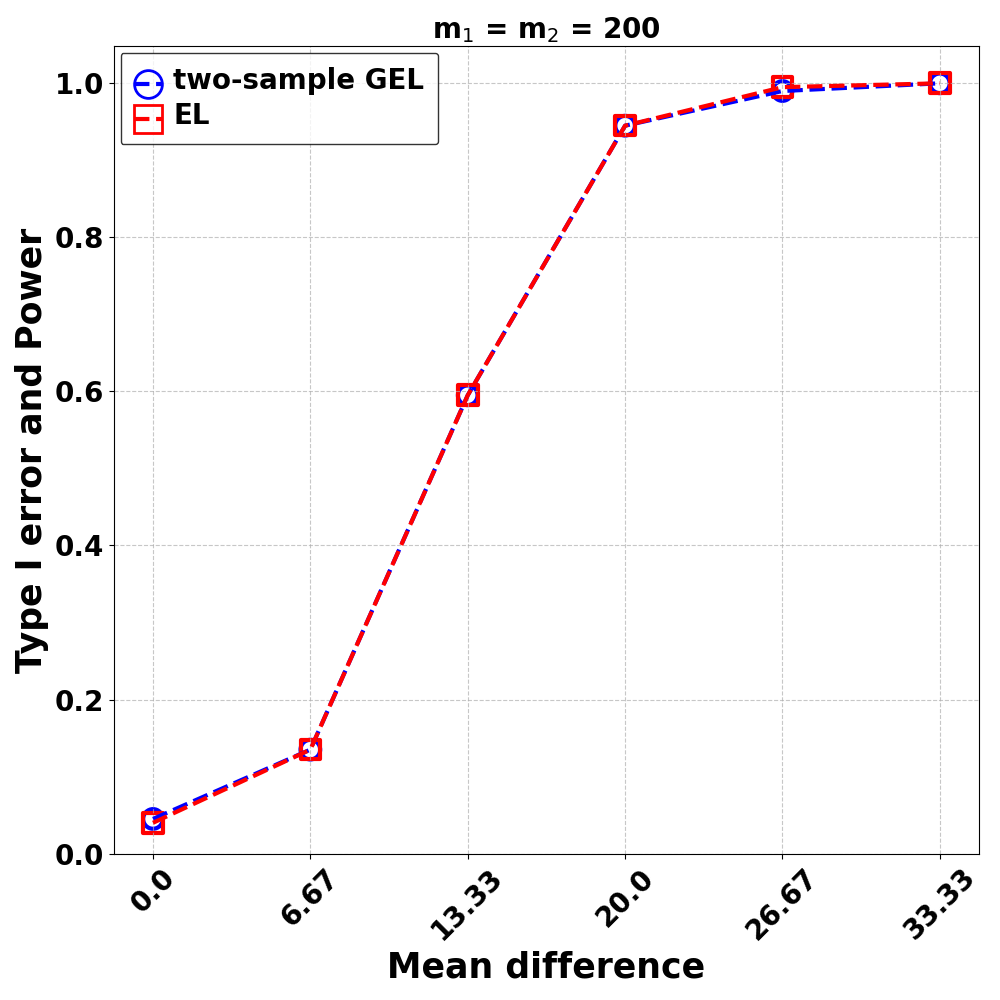}
  \end{subfigure}

  \vspace{4pt} 
  \begin{subfigure}[b]{0.45\linewidth}
    \centering
    \includegraphics[width=\linewidth]{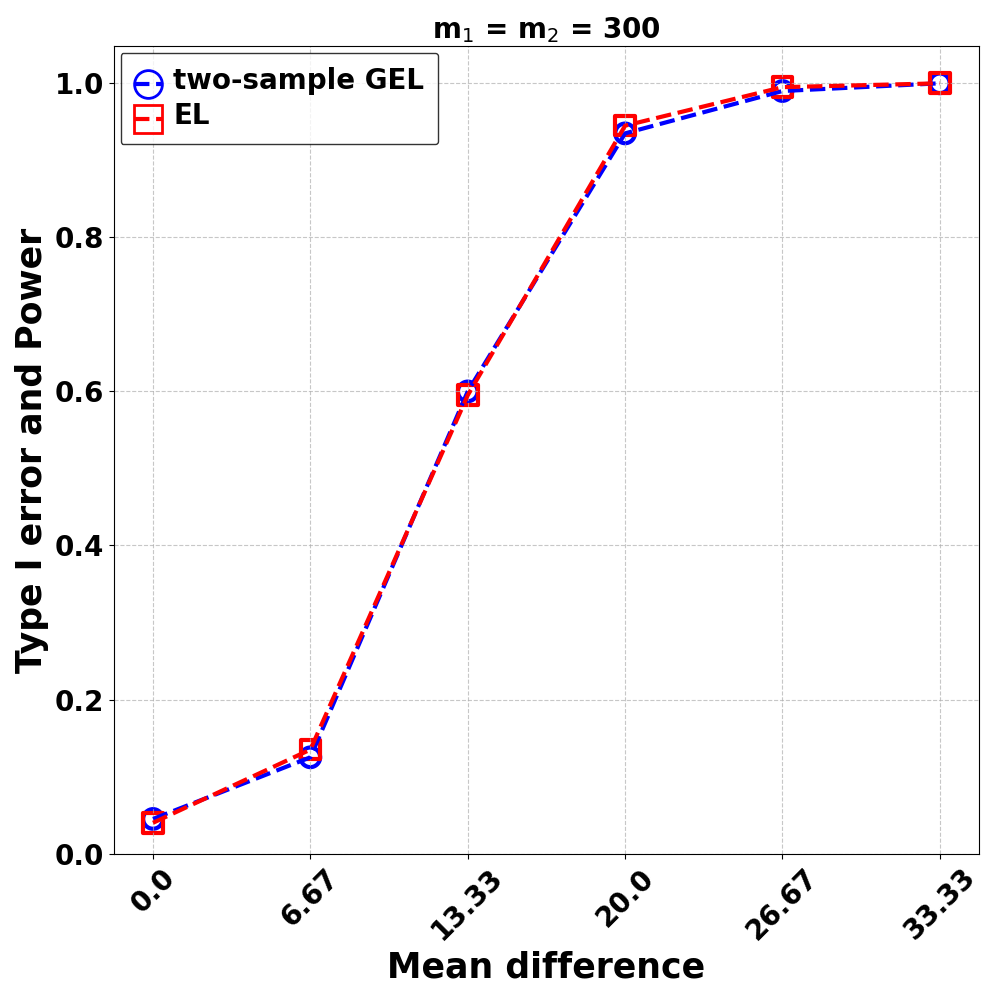}
  \end{subfigure}\hfill
  \begin{subfigure}[b]{0.45\linewidth}
    \centering
    \includegraphics[width=\linewidth]{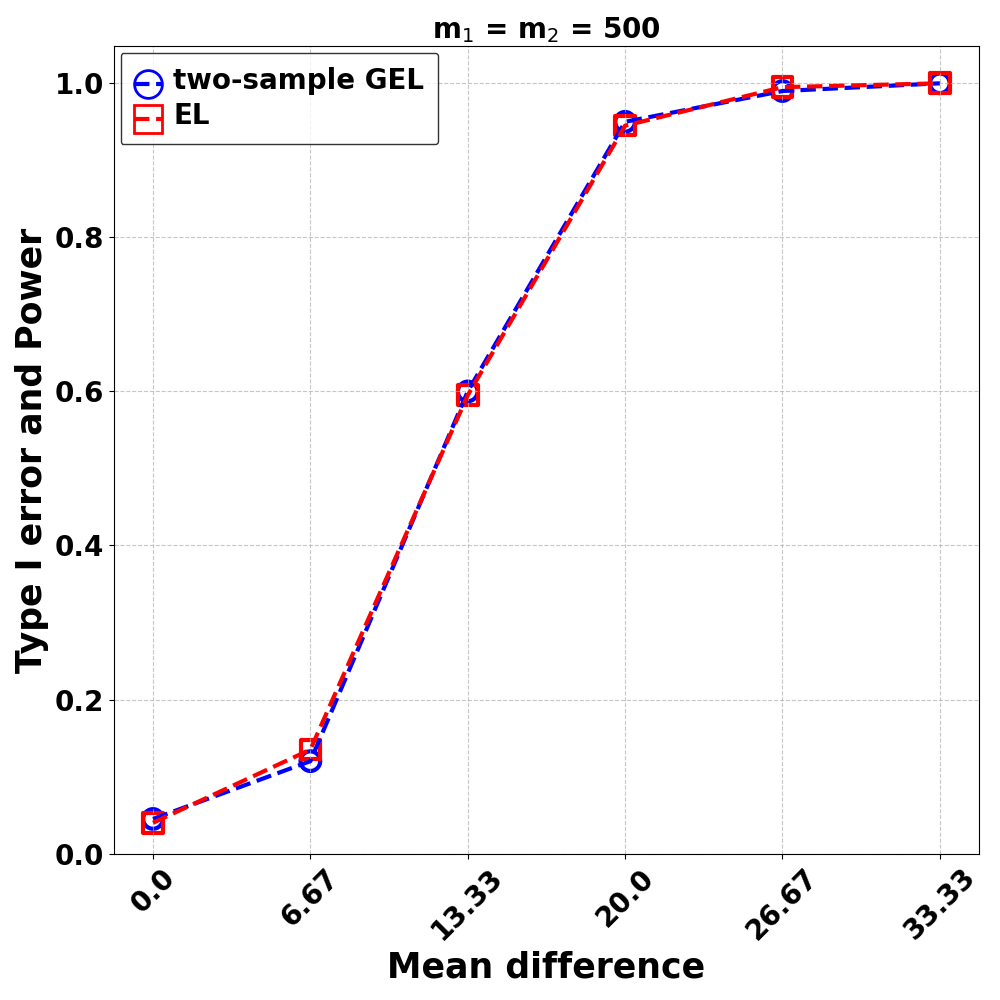}
  \end{subfigure}

  \caption{Comparison of EL and GEL estimators across different partition sizes.}
  \label{fig:two sample}
\end{figure}

Figure \ref{fig:two sample} presents the simulation results, comparing the standard two-sample EL method with our proposed two-sample GEL method.
It can be seen that the proposed GEL method possesses almost the same performance as the EL method, both under $H_0$ and $H_1$, demonstrating its effectiveness in detecting mean differences.
Furthermore, the two-sample GEL method offers substantial computational advantages. For example, with $m = 300\ (n=100)$, the average time for GEL is five times faster than the standard EL method.

\section{Real data analysis} \label{Real}

\subsection{Evaluation of GEL on the SOCR Height-Weight Dataset}

We conducted a comprehensive performance evaluation of our proposed GEL estimators using \href{https://wiki.socr.umich.edu/index.php/SOCR_Data_Dinov_020108_HeightsWeights}{\textcolor{blue}{SOCR Data Dinov 020108 HeightsWeights}} dataset. This dataset contains 25,000 anthropometric records documenting heights (in inches) and weights (in pounds) of 18-year-old individuals. We partition the first 20,000 observations as a training set and reserve the remaining 5,000 for testing. On the training data, we fit a linear regression model $W = \beta_{0} + \beta_{1} H$, where \( W \) denotes weight and \( H \) denotes height, by four approaches: ordinary least squares (OLS), EL, DCEL and GEL.

In the DCEL approach, the samples in the training set are randomly split into $k = 10, 100, 200, 500$ disjoint blocks. In contrast, for GEL, the same training samples are grouped into equal weighted subsets of size $m = 10, 100, 200, 500$, where the definitions of $m$ and $k$ are given in Example 1. Table \ref{tab:mspe_comparison} presents the mean squared prediction error (MSPE) on the test set across all methodologies. Numbers in parentheses denote the standard deviation of individual squared prediction errors $(\widehat{W}_{\text{predict}} - W_{\text{true}})^{2}$ across the 5,000 test cases, quantifying variability in predictive accuracy.

\begin{table}[htbp]
\centering
\caption{MSPE and average computation time for various methods on the test set (standard deviations in parentheses).}
\label{tab:mspe_comparison}
\setlength{\tabcolsep}{24pt}
\renewcommand{\arraystretch}{1.2}
\begin{tabular}{l c c}
\toprule
Method & MSPE & Time (sec.) \\
\midrule
OLS $(k=m=1)$   & 100.6264 (144.6296) & 0.0102 \\
EL  $(k=m=1)$   & 100.6264 (144.6296) & 12.8138\\
\midrule
DCEL $(k=10)$   & 100.6282 (144.6331) & 1.2900\\
GEL $(m=10, n=2000)$   & 100.6265 (144.6266) & 0.1253 \\
\midrule
DCEL $(k=50)$   & 100.6273 (144.6286) & 0.3029\\
GEL $(m=50, n=400)$   & 100.6239 (144.6278) & 0.0082\\
\midrule
DCEL $(k=100)$  & 100.6283 (144.6287) & 0.3671 \\
GEL $(m=100, n=200)$  & 100.6290 (144.6312) & 0.0038 \\
\midrule
DCEL $(k=200)$  & 100.6264 (144.6290) & 0.1262\\
GEL $(m=200, n=100)$  & 100.6260 (144.6143) & 0.0011 \\
\bottomrule
\end{tabular}
\end{table}

We can see that, the GEL method demonstrates compelling advantages across performance and efficiency metrics. It achieves predictive performance parity with both EL and OLS benchmarks, with MSPE values (100.6239--100.6290) essentially equivalent to centralized EL (100.6264) and OLS (100.6264). Computationally, GEL delivers dramatic efficiency improvements, achieving up to 11,600$\times$ acceleration over EL (0.0011s vs 12.8138s at $m=200$) and 115$\times$ faster execution than DCEL at comparable partition sizes (0.0011s vs 0.1262s). Remarkably, at $m=200$, GEL runs 9$\times$ faster than even highly optimized OLS implementations while maintaining equivalent accuracy. The method further demonstrates robustness to partitioning schemes, with negligible impact on predictive performance as evidenced by stable MSPE values (range: 0.0151) and standard deviations (range: 0.2177) across group sizes ($m=10$ to $m=200$).

Moreover, our simulations indicate that the GEL estimator is remarkably insensitive to the particular random split of the data. For instance, with \(m=10\), we performed 1000 independent random partitions of the training set, computed the GEL coefficient estimate for each partition, and averaged these 1000 estimates to obtain a final model. The resulting MSPE on the test set was 100.6264, which is virtually identical to the single-split value of 100.6265 reported in Table~\ref{tab:mspe_comparison}, demonstrating the robustness of GEL to sample division.

\subsection{Evaluation of two-sample GEL on the Gross Merchandise Volume Dataset}

The two-sample GEL method we proposed can be applied to controlled experiments within internet companies. Controlled experiments, commonly known as A/B tests, allow businesses to assess the impact of product changes by comparing key metrics (e.g., user engagement, revenue, conversion rates) between different treatment groups \cite{kohavi2020trustworthy,kohavi2023online, larsen2024statistical}. Controlled experiments have become a cornerstone of data-driven decision-making, especially in large-scale online platforms and technology companies.

Welch t-test (WT) is commonly used in traditional A/B tests \cite{welch1938significance, johari2017peeking}. However, when dealing with typical heavy-tailed big data commonly encountered in business contexts, such as Gross Merchandise Volume (GMV), we observed that the WT often fails to control Type I error in A/B tests platform. GMV is a key performance indicator (KPI) that measures the total sales volume of goods and services transacted through a company’s platform or marketplace \cite{prokhorova2020forecasting}. However, given the current limitations of optimization algorithms, the traditional two-sample EL method often fails when applied to large-scale datasets. The two-sample GEL method we proposed can address the aforementioned challenges.

At a certain’s A/B test platform, we randomly allocated 1$\%$ of weekly gross merchandise volume (GMV) traffic to both the experimental and control groups, each containing 204,000 observations. For two-sample GEL, we fixed the number of equal‐weight partitions at 1,000. With a significance level of 0.05, we conducted 1,000 times A/A tests using the WT method and the two-sample GEL method, producing 1,000 p-values. The WT method yielded a Type I error rate of 0.062, while two-sample GEL held it at 0.047. Although the average runtimes ( 0.1235 s and 0.1277 s) were comparable, two-sample GEL demonstrated superior control over the Type I error.

\section{Conclusion} \label{Con}

We have proposed the GEL method for nonparametric inference on massive datasets. It is shown that the proposed methods possess similar statistical properties to EL with much fewer parameters. Extensive simulations and real-data analyses demonstrate that GEL delivers competitive inferential accuracy while substantially shorter computation time. The selection of the group number $n$, also known as the effective sample size, depends on the complexity of the specific problem. Generally, we can recommend selecting $n$ to be of the order $100$.

Actually this paper reveals a parameters' dimensionality-reduction principle of EL. It is easy to extend this idea to other nonparametric likelihood settings, including Euclidean Likelihood \cite{owen1991empirical}, Bayesian empirical likelihood  \cite{lazar2003bayesian}, the bias-corrected empirical likelihood  \cite{zhu2006empirical}, the jackknife empirical likelihood  \cite{jing2009jackknife}, the adjusted empirical likelihood  \cite{chen2008adjusted}, the extended empirical likelihood  \cite{tsao2013extending}, the transformed empirical likelihood  \cite{jing2017transforming}, the mean empirical likelihood  \cite{liang2019mean}, and Bayesian penalized empirical likelihood \cite{chang2025bayesian}, We can also extend GEL to streaming data settings to support online inference \cite{aggarwal2007data, schifano2016online}.

\section*{Acknowledgement}
	
This work is supported by the National Natural Science Foundation of China (Grant No. 12571276).


\bibliographystyle{unsrt}  
\bibliography{references}  

\clearpage
\appendix

\section{Appendix}
We now present proofs of theorems in the order as they appeared in the paper. As a preliminary to the proof of the theorem, we state Lemma \ref{lem:3.1}.

\begin{lemma} \label{lem:3.1}
Assume that $\mathbb{E}\left[\v{g}(\bm{X}, \bm{\theta}_0)\v{g}^{'}(\bm{X}, \bm{\theta}_0)\right]$ is positive definite, $\partial \v{g}(\bm{x}, \bm{\theta})/\partial \bm{\theta}$ is continuous in a neighborhood of the true value $\bm{\theta}_0$, $\left\| \partial \v{g}(\bm{x}, \bm{\theta})/\partial \bm{\theta} \right\|$ and $\|\v{g}(\bm{x}, \bm{\theta})\|^3$ are bounded by some integrable function $H(\bm{x})$ in this neighborhood, and the rank of $\mathbb{E}[\partial \v{g}(\bm{X}, \bm{\theta}_0)/\partial \bm{\theta}]$ is $p$. Then, as $N \to \infty$, with probability $1$, $L_E(\bm{\theta})$ attains its minimum value at some point $\tilde{\bm{\theta}}$ in the interior of the ball $\| \bm{\theta} - \bm{\theta}_0 \| \leq N^{-1/3}$, and $\tilde{\bm{\theta}}$ and $\tilde{\bm{\lambda}} = \bm{\lambda}(\tilde{\bm{\theta}})$ satisfy
\[
    \v{Q}_{1n}(\tilde{\bm{\theta}}, \tilde{\bm{\lambda}}) = \v{0},
    \qquad
    \v{Q}_{2n}(\tilde{\bm{\theta}}, \tilde{\bm{\lambda}}) = \v{0},
\]
\end{lemma}
\noindent where
\begin{align}
    \v{Q}_{1n}(\bm{\theta}, \bm{\lambda}) &= \frac{1}{n} \sum_i \frac{\bar{\v{g}}(\bm{X}_i,\bm{\theta})}{1 + \bm{\lambda}^{'} \bar{\v{g}}(\bm{X}_i, \bm{\theta})} ,
    \label{eq:Q1_def} \\
    \v{Q}_{2n}(\bm{\theta}, \bm{\lambda}) &= \frac{1}{n} \sum_i \frac{1}{1 + \bm{\lambda}^{'} \bar{\v{g}}(\bm{X}_i, \bm{\theta})}
    \left( \frac{\partial \bar{\v{g}}(\bm{X}_i, \bm{\theta})}{\partial \bm{\theta}} \right)^{'} \bm{\lambda}.
\end{align}

\subsection{Proof of Lemma 1}
\begin{proof}
Define
\[
\v{G}_i(\bm{\theta})
:=\frac1m\sum_{j=1}^m \v{g}(\bm{X}_{ij},\bm{\theta}).
\]
Write
\(\bm{\theta}=\bm{\theta}_0+\bm{u}\,N^{-1/3}\) with \(\|\bm{u}\|=1\), and let \(\tilde{\bm{\lambda}}\) satisfy
\[
\frac1n\sum_{i=1}^n\frac{\v{G}_i(\bm{\theta})}{1+\tilde{\bm{\lambda}}^{'} \v{G}_i(\bm{\theta})}
=\v{0}.
\]
A Taylor expansion of $1/(1+\tilde{\bm{\lambda}}^{'} \v{G}_i(\bm{\theta}))$ around \(\tilde{\bm{\lambda}}=\v{0}\) gives, for some \(\xi_i\) on the line segment between \(0\) and \(\tilde{\bm{\lambda}}\)\,,
\[
\frac{\v{G}_i(\bm{\theta})}{1+\tilde{\bm{\lambda}}^{'} \v{G}_i(\bm{\theta})}
=\v{G}_i(\bm{\theta})\bigl(1-\xi_i\,\tilde{\bm{\lambda}}^{'} \v{G}_i(\bm{\theta})\bigr)
+o_p\bigl((m/n)^{1/3}\bigr).
\]
Hence
\begin{align*}
\v{0}
&=\frac1n\sum_{i=1}^n\v{G}_i(\bm{\theta})\bigl(1-\xi_i\,\tilde{\bm{\lambda}}^{'} \v{G}_i(\bm{\theta})\bigr)
+o_p\bigl((m/n)^{1/3}\bigr)\\
&=\frac1n\sum_{i=1}^n\v{G}_i(\bm{\theta})
-\frac1n\sum_{i=1}^n\v{G}_i(\bm{\theta})\v{G}_i(\bm{\theta})^{'}\,\tilde{\bm{\lambda}}
+o_p\bigl((m/n)^{1/3}\bigr).
\end{align*}
Rearranging and inverting the empirical second‐moment matrix,
\[
\tilde{\bm{\lambda}}
=\Bigl(\frac{1}{n}\sum_{i=1}^n\v{G}_i(\bm{\theta})\v{G}_i(\bm{\theta})^{'}\Bigr)^{-1}
\frac{1}{n}\sum_{i=1}^n\v{G}_i(\bm{\theta})
+o_p\bigl((m/n)^{1/3}\bigr)
=O_p\bigl((m/n)^{1/3}\bigr),
\]
uniformly for \(\bm{\theta}\) in the neighborhood \(\|\bm{\theta}-\bm{\theta}_0\|\le N^{-1/3}\).
Now for the log-likelihood
\begin{align*}
\frac{-2\log R_G(\bm{\theta})}{m} &= \sum_{i=1}^{n} \log (1 + \tilde{\bm{\lambda}}^{'} \v{G}_i(\bm{\theta})) = \sum_{i=1}^{n} \left( \tilde{\bm{\lambda}}^{'} \v{G}_i(\bm{\theta}) - \frac{1}{2} (\tilde{\bm{\lambda}}^{'} \v{G}_i(\bm{\theta}))^2 \right) + o_p((n/m)^{1/3})\\
&= \frac{n}{2} \left( \frac{1}{n} \sum_{i=1}^{n} \v{G}_i(\bm{\theta}) \right)^{'} \left( \frac{1}{n} \sum_{i=1}^{n} \v{G}_i(\bm{\theta}) \v{G}_i(\bm{\theta})^{'} \right)^{-1} \left( \frac{1}{n} \sum_{i=1}^{n} \v{G}_i(\bm{\theta}) \right) + o_p((n/m)^{1/3}).
\end{align*}
Expanding around $\bm{\theta}_0$, we have
\begin{align*}
\frac{-2\log R_G(\bm{\theta})}{m}&= \frac{n}{2} \left( \frac{1}{n} \sum_{i=1}^{n} \v{G}_i(\bm{\theta}) + \frac{1}{n} \sum_{i=1}^{n} \frac{\partial \v{G}_i(\bm{\theta})}{\partial \bm{\theta}} \bm{u} N^{-1/3} \right)^{'} \left( \frac{1}{n} \sum_{i=1}^{n} \v{G}_i(\bm{\theta}) \v{G}_i(\bm{\theta})^{'} \right)^{-1}\\
&\times \left( \frac{1}{n} \sum_{i=1}^{n} \v{G}_i(\bm{\theta}) + \frac{1}{n} \sum_{i=1}^{n} \frac{\partial \v{G}_i(\bm{\theta})}{\partial \bm{\theta}} \bm{u} N^{-1/3} \right) + o_p((n/m)^{1/3})\\
&= \frac{n}{2} \left\{ O_p(N^{-1/2}(\log \log N)^{1/2}) + \E\left( \frac{\partial \v{G}(\bm{\theta}_0)}{\partial \bm{\theta}} \right) \bm{u} N^{-1/3} \right\}^{'} \E\left( \v{G}(\bm{\theta}_0) \v{G}^{'}(\bm{\theta}_0) \right)^{-1}\\
&\times \left\{ O_p(N^{-1/2}(\log \log N)^{1/2}) + \E\left( \frac{\partial \v{G}(\bm{\theta}_0)}{\partial \bm{\theta}} \right) \bm{u} N^{-1/3} \right\} + o_p((n/\log n)^{1/3})\\
&\geq \frac{n}{2} \cdot (mc - m\epsilon) \cdot N^{-2/3} = (c - \epsilon) \cdot N^{1/3} \quad \text{as}\quad N \to \infty .
\end{align*}
Note that
\[
\frac{\partial \v{G}_i(\bm{\theta})}{\partial \bm{\theta}} = \frac{\partial}{\partial \bm{\theta}} \left( \frac{1}{m} \sum_{j=1}^m \v{g}(\bm{X}_{ij}, \bm{\theta}) \right) = \frac{1}{m} \sum_{j=1}^m \frac{\partial \v{g}(\bm{X}_{ij}, \bm{\theta})}{\partial \bm{\theta}}.
\]
Therefore, at $\bm{\theta}_0$, we have
\[
\E\left( \frac{\partial \v{G}(\bm{\theta}_0)}{\partial \bm{\theta}} \right) = \frac{1}{m} \sum_{j=1}^m \E\left( \frac{\partial \v{g}(\bm{X}_{ij}, \bm{\theta}_0)}{\partial \bm{\theta}} \right) = \E\left( \frac{\partial \v{g}(\bm{X}, \bm{\theta}_0)}{\partial \bm{\theta}} \right)
\]
and
\[
\E\left( \v{G}(\bm{\theta}_0) \v{G}^{'}(\bm{\theta}_0) \right) = \frac{1}{m} \E\left( \v{g}(\bm{X}, \bm{\theta}_0) \v{g}^{'}(\bm{X}, \bm{\theta}_0) \right).
\]
Let $c$ be the smallest eigenvalue of
\[
\E\left( \frac{\partial \v{g}(\bm{X}, \bm{\theta}_0)}{\partial \bm{\theta}} \right)^{'} \left\{\E\left( \v{g}(\bm{X}, \bm{\theta}_0) \v{g}^{'}(\bm{X}, \bm{\theta}_0) \right)\right\}^{-1} \E\left( \frac{\partial \v{g}(\bm{X}, \bm{\theta}_0)}{\partial \bm{\theta}} \right),
\]
then the smallest eigenvalue of
\[
\E\left( \frac{\partial\v{G}(\bm{\theta}_0)}{\partial \bm{\theta}} \right)^{'} \E\left( \v{G}(\bm{\theta}_0) \v{G}^{'}(\bm{\theta}_0) \right)^{-1} \E\left( \frac{\partial \v{G}(\bm{\theta}_0)}{\partial \bm{\theta}} \right)
\]
is $mc$.
Similarly, at $\bm{\theta}_0$, we have
\begin{align*}
\frac{-2\log R_G(\bm{\theta}_0)}{m} &= \frac{n}{2} \left( \frac{1}{n} \sum_{i=1}^n \v{G}_i(\bm{\theta}_0) \right)^{'} \left( \frac{1}{n} \sum_{i=1}^n \v{G}_i(\bm{\theta}_0) \v{G}_i^{'}(\bm{\theta}_0) \right)^{-1} \left( \frac{1}{n} \sum_{i=1}^n \v{G}_i(\bm{\theta}_0) \right) + o_p(1)\\
&= \frac{N}{2} \left(\frac{1}{N} \sum_{i=1}^N \v{g}(\bm{X}_i, \bm{\theta}_0) \right)^{'}  \left( \frac{1}{N} \sum_{i=1}^N \v{g}(\bm{X}_i, \bm{\theta}_0) \v{g}^{'}(\bm{X}_i, \bm{\theta}_0) \right)^{-1}  \left( \frac{1}{N} \sum_{i=1}^N \v{g}(\bm{X}_i, \bm{\theta}_0) \right) + o_p(1)\\
&= O_p(\log \log N).
\end{align*}
From $-2\log R_G(\bm{\theta}) > -2\log R_G(\bm{\theta}_0)$ for $\|\bm{\theta} - \bm{\theta}_0\| = N^{-1/3}$, we see that $-2\log R_G(\bm{\theta})$ is larger on the sphere than at the center. By continuity, $R_G(\bm{\theta})$ attains a minimum at some interior point of the ball, and $\hat{\bm{\theta}}$ satisfies
\begin{align*}
\frac{-2\partial\log R_G(\bm{\theta})}{\partial \bm{\theta}} \bigg|_{\bm{\theta}=\hat{\bm{\theta}}} &= \sum_{i=1}^{n}\frac{\frac{\partial \tilde{\bm{\lambda}}(\bm{\theta})}{\partial\bm{\theta}} \v{G}_i(\bm{\theta})+\left(\frac{\partial \v{G}_{i}(\bm{\theta})}{\partial \bm{\theta}}\right)^{'}\tilde{\bm{\lambda}}(\bm{\theta})}{1+\tilde{\bm{\lambda}}^{'}(\bm{\theta})\v{G}_{i}(\bm{\theta})}\bigg|_{\bm{\theta}=\hat{\bm{\theta}}} \\
&= \sum_{i=1}^{n}\frac{1}{1+\tilde{\bm{\lambda}}^{'}(\bm{\theta})\v{G}_{i}(\bm{\theta})}\left(\frac{\partial \v{G}_{i}(\bm{\theta})}{\partial \bm{\theta}}\right)^{'}\tilde{\bm{\lambda}}(\bm{\theta})\bigg|_{\bm{\theta}=\hat{\bm{\theta}}}\\
&=\v{0}.
\end{align*}
Thus
\[
\v{Q}_{1n}(\tilde{\bm{\theta}},\tilde{\bm{\lambda}})=\frac{1}{n}\sum_{i=1}^{n}\frac{\v{G}_{i}(\tilde{\bm{\theta}})}{1+\tilde{\bm{\lambda}}^{'} \v{G}_{i}(\tilde{\bm{\theta}})} = \v{0},
\]
\[
\v{Q}(\tilde{\bm{\theta}},\tilde{\bm{\lambda}})=\frac{1}{n}\sum_{i=1}^{n}\frac{1}{1+\tilde{\bm{\lambda}}^{'} \v{G}_{i}(\tilde{\bm{\theta}})}\left(\frac{\partial \v{G}_{i}(\tilde{\bm{\theta}})}{\partial \bm{\theta}}\right)^{'}\tilde{\bm{\lambda}} = \v{0},
\]
where
\[
\v{Q}(\bm{\theta},\bm{\lambda})=\frac{1}{n}\sum_{i=1}^{n}\frac{\v{G}_{i}(\bm{\theta})}{1+\bm{\lambda}^{'} \v{G}_{i}(\bm{\theta})},
\]
\[
\v{Q}(\bm{\theta},\bm{\lambda})=\frac{1}{n}\sum_{i=1}^{n}\frac{1}{1+\bm{\lambda}^{'} \v{G}_{i}(\bm{\theta})}\left(\frac{\partial \v{G}_{i}(\bm{\theta})}{\partial \bm{\theta}}\right)^{'} \bm{\lambda}.
\]
\end{proof}

\subsection{Proof of Theorem 1}
\begin{proof} Taking derivatives about $\bm{\theta}$ and $\bm{\lambda}^{'}$, we have
\[
\frac{\partial \v{Q}_{1n}(\bm{\theta}, \bm{0})}{\partial \bm{\theta}} = \frac{1}{n} \sum_{i=1}^{n} \frac{\partial \v{G}_i(\bm{\theta})}{\partial \bm{\theta}}  = \frac{1}{N} \sum_{i=1}^{N} \frac{\partial \v{g}(\bm{X}_i, \bm{\theta})}{\partial \bm{\theta}}
\]
\[
\frac{\partial \v{Q}_{1n}(\bm{\theta}, 0)}{\partial \bm{\lambda}^{'}} = -\frac{1}{n} \sum_{i=1}^{n} \v{G}_i(\bm{\theta}) \v{G}_i(\bm{\theta})^{'}
\]
\[
\frac{\partial \v{Q}_{2n}(\bm{\theta}, \v{0})}{\partial \bm{\theta}} = \v{0}
\]
\[
\frac{\partial \v{Q}_{2n}(\bm{\theta}, \v{0})}{\partial \bm{\lambda}^{'}} = \frac{1}{n} \sum_{i=1}^{n} \left( \frac{\partial \v{G}_i(\bm{\theta})}{\partial \bm{\theta}} \right)^{'}
\]

Expanding $\v{Q}_{1n}(\hat{\bm{\theta}}, \tilde{\bm{\lambda}})$ and $\v{Q}_{2n}(\hat{\bm{\theta}}, \tilde{\bm{\lambda}})$ at $(\bm{\theta}_0, \v{0})$ by the conditions of the Theorem 1 and Lemma 1, we have
\begin{align*}
\v{0} &= \v{Q}_{1n}(\hat{\bm{\theta}}, \tilde{\bm{\lambda}})\\
&= \v{Q}_{1n}(\bm{\theta}_0, \v{0}) + \frac{\partial \v{Q}_{1n}(\bm{\theta}_0, \v{0})}{\partial \bm{\theta}} (\tilde{\bm{\theta}} - \bm{\theta}_0) + \frac{\partial \v{Q}_{1n}(\bm{\theta}_0, \v{0})}{\partial \bm{\lambda}^{'}} ( \tilde{\bm{\lambda}}-\v{0}) + r_N
\end{align*}
\begin{align*}
\v{0} &= \v{Q}_{2n}(\hat{\bm{\theta}}, \tilde{\bm{\lambda}}) \\
&= \v{Q}_{2n}(\bm{\theta}_0, \v{0}) + \frac{\partial \v{Q}_{2n}(\bm{\theta}_0, \v{0})}{\partial \bm{\theta}} (\tilde{\bm{\theta}} - \bm{\theta}_0) + \frac{\partial \v{Q}_{2n}(\bm{\theta}_0, \v{0})}{\partial \bm{\lambda}^{'}} ( \tilde{\bm{\lambda}}-\v{0}) +  r_N
\end{align*}
where $ r_N= o_p(\|\hat{\bm{\theta}} - \bm{\theta}_0\| + \|\tilde{\bm{\lambda}}\|)$. We have
\[
\begin{pmatrix}
\tilde{\bm{\lambda}}\\
\hat{\bm{\theta}} - \bm{\theta}_0
\end{pmatrix}
= \bm{S}_N^{-1} \begin{pmatrix}
-\v{Q}_{1n}(\bm{\theta}_0, \bm{0})+ r_N \\
r_N
\end{pmatrix},
\]
where
\[
\bm{S}_N = \begin{pmatrix}
\frac{\partial \v{Q}_{1n}}{\partial \bm{\lambda}^{'}} & \frac{\partial \v{Q}_{1n}}{\partial \bm{\theta}} \\
\frac{\partial \v{Q}_{2n}}{\partial \bm{\lambda}^{'}} & \v{0}
\end{pmatrix}_{(\bm{\theta}_0,\v{0})}
\to \begin{pmatrix}
\bm{S}_{11} & \bm{S}_{12} \\
\bm{S}_{21} &\v{0}
\end{pmatrix} = \begin{pmatrix}
-\E(\v{G} \v{G}^{'}) &  \E(\frac{\partial \v{G}}{\partial \bm{\theta}})\\
\E(\frac{\partial \v{G}}{\partial \bm{\theta}})^{'} & \v{0}
\end{pmatrix}.
\]
Let $\bm{S}_{21}^*=-(\bm{S}_{21}\bm{S}_{11}^{-1}\bm{S}_{12})^{-1}=\left\{ \E(\frac{\partial \v{G}}{\partial \bm{\theta}})^{'}\E(\v{G} \v{G}^{'})^{-1}\E(\frac{\partial \v{G}}{\partial \bm{\theta}}) \right\}^{-1}$, we can get
\[
\hat{\bm{\theta}} - \bm{\theta}_0 = \bm{S}_{21}^*\bm{S}_{21}\bm{S}_{11}^{-1}\v{Q}_{1n}(\bm{\theta}_0, \bm{0}) - \bm{S}_{21}^*\bm{S}_{21}\bm{S}_{11}^{-1}r_N + \bm{S}_{21}^*r_N.
\]
Thus
\[
\sqrt{N} (\hat{\bm{\theta}} - \bm{\theta}_0) = \bm{S}_{21}^* \bm{S}_{21} \bm{S}_{11}^{-1} \sqrt{N} \v{Q}_{1n}(\bm{\theta}_0, \v{0}) + o_p(1) \dto \mathcal{N}(\v{0}, \frac{\bm{V}}{m}),
\]
where $\bm{V}$ is given by (\ref{V_}).
\end{proof}

\subsection{Proof of Theorem 2}
\begin{proof}
When $r=p$, we can get
\[
q_i = \frac{1}{N\{1 + \tilde{\bm{\lambda}}^{'}\bar{\v{g}}(\bm{X}_{i},\bm{\theta})\}}.
\]
From the constraint \(\sum_{i=1}^n q_i \sum_{j=1}^m \v{g}(\bm{X}_{ij},\bm{\theta}) = \v{0}\), we have
\[
\frac{1}{n} \sum_{i=1}^n \frac{\bar{\v{g}}(\bm{X}_{i},\bm{\theta})}{1 + \tilde{\bm{\lambda}}^{'} \bar{\v{g}}(\bm{X}_{i},\bm{\theta})} = \v{0}.
\]
Let
\[
Z_N^* = \max_{\substack{1 \leq i \leq n \\ 1 \leq j \leq m }}
 \|\v{g}(\bm{X}_{ij},\bm{\theta})\|,\quad
\bar{Z}_n^* = \max_{1 \leq i \leq n} \|\bar{\v{g}}(\bm{X}_{i},\bm{\theta})\|,
\]
and we have
\[
Z_N^* = o_p (N^{\frac{1}{2}}).
\]
Since \(\var(\sqrt{m}\bar{\v{g}}(\bm{X}_{i},\bm{\theta}))  < \infty\), we have
\[
\sum_{i=1}^n \Pr(\|\sqrt{m}\bar{\v{g}}(\bm{X}_{i},\bm{\theta})\| > n) < \infty.
\]
By the Borel-Cantelli lemma, the event \(\|\sqrt{m}\bar{\v{g}}(\bm{X}_{n},\bm{\theta})\| > \sqrt{n}\) occurs only for finitely many \(n\), almost surely. This implies that for all sufficiently large \(n\)
\[
\sqrt{m} \bar{Z}_n^* \leq \sqrt{n} \quad \text{almost surely}.
\]
That is, for \(\forall \epsilon > 0\)
\[
\sqrt{m} \bar{Z}_n^* \leq \epsilon n^{\frac{1}{2}} \quad \text{almost surely for large } n.
\]
Therefore,
\begin{align*}
&\Rightarrow \limsup_{n \to \infty} \bar{Z}_n^* \sqrt{m} \cdot \frac{1}{\sqrt{n}} \leq \epsilon \\
&\Rightarrow \bar{Z}_n^* = o_p\left(\sqrt{\frac{n}{m}}\right).
\end{align*}

Let $\tilde{\bm{\lambda}}=\|\tilde{\bm{\lambda}}\|\bm{I}$ where $\bm{I}$ is a unit vector. Define $Y_i := \tilde{\bm{\lambda}}^{'} \bar{\v{g}}(\bm{X}_{i},\bm{\theta})$ and
\[
A(\tilde{\bm{\lambda}}) := \frac{1}{n} \sum_{i=1}^n \frac{\bar{\v{g}}(\bm{X}_{i},\bm{\theta})}{1 + \tilde{\bm{\lambda}}^{'} \bar{\v{g}}(\bm{X}_{i},\bm{\theta})} = 0.
\]
Due to $1/(1 + Y_i) = 1 - Y_i/(1 + Y_i)$ and substituting into $\bm{I}^{'} A(\tilde{\bm{\lambda}}) = 0$, we can get
\[
\| \tilde{\bm{\lambda}} \| \bm{I}^{'} \tilde{\bm{S}} \bm{I} = \bm{I}^{'} \bar{\v{g}}(\bm{X},\bm{\theta}),
\]
where
\[
\tilde{\bm{S}} = \frac{1}{n} \sum_{i=1}^n \frac{\bar{\v{g}}(\bm{X}_{i},\bm{\theta})\bar{\v{g}}(\bm{X}_{i},\bm{\theta})^{'}}{1 + Y_i}
\]
and $\bar{\v{g}}(\bm{X},\bm{\theta})=\sum_{i=1}^{n}\bar{\v{g}}(\bm{X}_{i},\bm{\theta})/n$.
Since \(1 + Y_i > 0\), we have
\[
\| \tilde{\bm{\lambda}} \| \bm{I}^{'} \bm{S} \bm{I} \leq \| \tilde{\bm{\lambda}} \| \bm{I}^{'} \tilde{\bm{S}} \bm{I} (1 + \max_i |Y_i|)
\]
\[
\leq \| \tilde{\bm{\lambda}} \| \bm{I}^{'} \tilde{\bm{S}} \bm{I} (1 + \| \tilde{\bm{\lambda}} \| \bar{Z}_n^*),
\]
where $\bm{S}= (1/n)\sum_{i=1}^{n}\bar{\v{g}}(\bm{X}_{i},\bm{\theta})\bar{\v{g}}(\bm{X}_{i},\bm{\theta})^{'}$.
Thus
\[
\| \tilde{\bm{\lambda}} \| \left(\bm{I}^{'} \bm{S} \bm{I} - \bar{Z}_n^* \bm{I}^{'} \bar{\v{g}}(\bm{X},\bm{\theta})\right) \leq \bm{I}^{'} \bar{\v{g}}(\v{X},\bm{\theta}).
\]
Given that $\bm{I}^{'} \bar{\v{g}}(\bm{X},\bm{\theta}) = O_p (N^{-\frac{1}{2}})$, we can get
\[
\bar{Z}_n^*\bm{I}^{'} \bar{\v{g}}(\bm{X},\bm{\theta}) = o_p\left(\sqrt{\frac{n}{m}}.\sqrt{\frac{1}{N}}\right)=o_p\left(\frac{1}{m}\right),
\]
\[
\frac{\sigma_1}{m}+o_p\left(\frac{1}{m}\right) \geq \bm{I}^{'} \bm{S} \bm{I} \geq \frac{\sigma_p}{m}+o_p\left(\frac{1}{m}\right),
\]
where $\sigma_1$ denotes the minimum eigenvalue of $\var(\v{g}(\bm{X},\bm{\theta}))$, and $\sigma_p$ denotes the maximum eigenvalue of $\var(\v{g}(\bm{X},\bm{\theta}))$. Therefore, we have $\bm{I}^{'} \bm{S} \bm{I}=O_p\left(1/m\right)$. Next, we have
\[
\| \tilde{\bm{\lambda}} \| \left(O_p\left(\frac{1}{m}\right) + o_p\left(\frac{1}{m}\right)\right) = O_p(N^{-\frac{1}{2}})
\]
\[
\Rightarrow \| \tilde{\bm{\lambda}} \| = O_p\left(\sqrt{\frac{m}{n}}\right).
\]
Therefore,
\[
\max_{1 \leq i \leq n} |Y_i| = \max_{1 \leq i \leq n} \tilde{\bm{\lambda}}^{'} \bar{\v{g}}(\bm{X}_{i},\bm{\theta}) = O_p\left(\sqrt{\frac{m}{n}}\right) \cdot o_p\left(\sqrt{\frac{n}{m}}\right) = o_p(1).
\]
From $A(\tilde{\bm{\lambda}})$, we can get
\begin{align*}
\v{0} &= \frac{1}{n} \sum_{i=1}^{n} \bar{\v{g}}(\bm{X}_{i},\bm{\theta})\left(1 - Y_i + \frac{Y_i^2}{1 + Y_i}\right)\\
&= \bar{\v{g}}(\bm{X},\bm{\theta}) - \bm{S}\tilde{\bm{\lambda}} + \frac{1}{n} \sum_{i=1}^{n} \frac{\bar{\v{g}}(\bm{X}_{i},\bm{\theta})Y_i^2}{1 + Y_i},
\end{align*}
where
\[
 \frac{1}{n} \sum_{i=1}^{n} \frac{\bar{\v{g}}(\bm{X}_{i},\bm{\theta})Y_i^2}{1 + Y_i}  = \frac{1}{n} \sum_{i=1}^{n} \|\bar{\v{g}}(\bm{X}_{i},\bm{\theta})\|^3 \| \tilde{\bm{\lambda}} \|^2 |1 + Y_i|^{-1}.
\]
Since
\[
\frac{1}{n} \sum_{i=1}^{n} \|\bar{\v{g}}(\bm{X}_{i},\bm{\theta})\|^3  \leq \frac{\bar{Z}_n^*}{n}\sum_{i=1}^{n} \|\bar{\v{g}}(\bm{X}_{i},\bm{\theta})\|^2
\]
and \(\var(\bar{\v{g}}(\bm{X}_{i},\bm{\theta})) = \var(\v{g}(\bm{X},\bm{\theta}))/m = O(1/m)\), we have
\[
\frac{\bar{Z}_n^*}{n}\sum_{i=1}^{n} \|\bar{\v{g}}(\bm{X}_{i},\bm{\theta})\|^2 = o_p\left(\sqrt{\frac{n}{m}}\right)\cdot O_p\left( \frac{1}{m}\right) = o_p\left(\frac{n^{1/2}}{m^{3/2}}\right)
\]
\[
\Rightarrow\frac{1}{n} \sum_{i=1}^{n} \|\bar{\v{g}}(\bm{X}_{i},\bm{\theta})\|^3 =  o_p\left(\frac{n^{1/2}}{m^{3/2}}\right).
\]
\[
\Rightarrow\frac{1}{n} \sum_{i=1}^{n} \frac{\bar{\v{g}}(\bm{X}_{i},\bm{\theta})Y_i^2}{1 + Y_i} = o_p\left(\frac{n^{1/2}}{m^{3/2}}\right) \cdot O_p\left(\frac{m}{n}\right) \cdot O_p(1) = o_p\left(\frac{1}{\sqrt{N}}\right).
\]
Therefore, we can get
\[
\tilde{\bm{\lambda}} = \bm{S}^{-1}\bar{\v{g}}(\bm{X},\bm{\theta}) + \bm\beta,
\]
where $\bm\beta = o_p(1/\sqrt{N})$.

Using the Taylor expansion for logarithm, we have
\[
\log(1 + Y_i) = Y_i - \frac{1}{2} Y_i^2 + \eta_i,
\]
for some finite $B > 0$, we have
\[
\Pr\left( |\eta_i| \leq B|Y_i|^3, 1\leq i \leq n\right) \to 1, \quad \text{as} \quad n \to \infty.
\]
Thus the log-likelihood ratio is
\begin{align*}
\frac{-2\log R_G(\bm{\theta}_0)}{m}
&= 2 \sum_{i=1}^{n} \log \{1 + \tilde{\bm{\lambda}}^{'} \bar{\v{g}}(\bm{X}_{i},\bm{\theta})\} = 2 \sum_{i=1}^{n} \log \{1 + Y_i\}\\
&= 2 \sum_{i=1}^{n} Y_i - \sum_{i=1}^{n} Y_i^2 + 2 \sum_{i=1}^{n} \eta_i\\
&= 2n\tilde{\bm{\lambda}}^{'}\bar{\v{g}}(\bm{X},\bm{\theta}) - n\tilde{\bm{\lambda}}^{'} \bm{S} \tilde{\bm{\lambda}} + 2 \sum_{i=1}^n \eta_i\\
&= n\bar{\v{g}}(\bm{X},\bm{\theta})^{'} \bm{S}^{-1}\bar{\v{g}}(\bm{X},\bm{\theta}) - n\bm\beta^{'} \bm{S} \bm\beta + 2 \sum_{i=1}^n \eta_i.
\end{align*}
The first term converges to \(\chi_p^2\). The remainder terms satisfy:
\begin{align*}
n\bm\beta^{'} \bm{S} \bm\beta &= n \cdot o_p\left(\frac{1}{\sqrt{N}}\right) \cdot O_p\left(\frac{1}{m}\right) \cdot o_p\left(\frac{1}{\sqrt{N}}\right) \\
&= o_p\left(\frac{1}{m^2}\right)
\end{align*}
and
\begin{align*}
\left| \sum_{i=1}^n \eta_i \right| &\leq B \|\tilde{\bm{\lambda}}\|^3 \sum_{i=1}^n \|\bar{\v{g}}(\bm{X}_{i},\bm{\theta})\|^3\\
&= O_p\left(\frac{m^{3/2}}{n^{3/2}}\right) \cdot n \cdot o_p\left(\frac{n^{1/2}}{m^{3/2}}\right)\\
&= o_p(1)
\end{align*}
Thus $-2\log R_G(\bm{\theta}_0)/m \dto \chi_p^2$.

When $r \textgreater p$,
\[
\frac{-2\log R_G(\bm{\theta}_0)}{m} = \frac{2}{m} \left\{ m \sum_{i=1}^n \log \left(1 + \tilde{\bm{\lambda}}^{'} \bm{G}_i (\bm{\theta}_0)\right) - m \sum_{i=1}^n \log \left( 1 + \tilde{\bm{\lambda}}^{'} \bm{G}_i (\hat{\bm{\theta}}) \right) \right\}.
\]
Note that
\[
-2\log R_G(\hat{\bm{\theta}}, \tilde{\bm{\lambda}}) = \sum_{i=1}^n \log \left(1 + \tilde{\bm{\lambda}}^{'} \bm{G}_i (\hat{\bm{\theta}})\right) = -\frac{n}{2} \bm{Q}_{1n}(\bm{\theta}_0, \bm{0})^{'} \bm{B} \bm{Q}_{1n}(\bm{\theta}_0, \v{0}) + o_p(1),
\]
where $\bm{B} = \bm{S}_{11}^{-1} \left(\bm{I} + \bm{S}_{12} \bm{S}_{21}^{*} \bm{S}_{21} \bm{S}_{11}^{-1} \right)$. Also under $H_0$,
\[
\frac{1}{n}\sum_{i=1}^{n}\frac{1}{1 + \tilde{\bm{\lambda}}_0^{'} \bm{G}_i (\bm{\theta}_0)} \bm{G}_i (\bm{\theta}_0) = \v{0} \quad \Rightarrow \quad \tilde{\bm{\lambda}}_0 = -\bm{S}_{11}^{-1}\bm{Q}_{1n}(\bm{\theta}_0,\v{0}) + o_p(1)
\]
and
\[
\sum_{i=1}^n \log \left(1 + \tilde{\bm{\lambda}}_0^{'} \bm{G}_i (\bm{\theta}_0)\right) = -\frac{n}{2} \bm{Q}_{1n}^{'}(\bm{\theta}_0, \v{0}) \bm{S}_{11}^{-1} \bm{Q}_{1n}(\bm{\theta}_0, \v{0}) + o_p(1).
\]
Thus
\begin{align*}
\frac{-2\log R_G(\bm{\theta}_0)}{m} &= n \bm{Q}_{1n}^{'}(\bm{\theta}_0, \v{0}) (\bm{B} - \bm{S}_{11}^{-1}) \bm{Q}_{1n}(\bm{\theta}_0, \v{0}) + o_p(1)\\
&= n \bm{Q}_{1n}^{'}(\bm{\theta}_0, \v{0}) \bm{S}_{11}^{-1} \bm{S}_{12} \bm{S}_{21}^{*}\bm{S}_{21} \bm{S}_{11}^{-1} \bm{Q}_{1n}(\bm{\theta}_0, \v{0}) + o_p(1)\\
&=\left[(-\bm{S}_{11})^{-\frac{1}{2}}\sqrt{n}\bm{Q}_{1n}(\bm{\theta}_0, \v{0})\right]^{'}\left[(-\bm{S}_{11})^{-\frac{1}{2}} \bm{S}_{12} \bm{S}_{21}^{*}\bm{S}_{21}(-\bm{S}_{11})^{-\frac{1}{2}}\right]\\
& \cdot \left[(-\bm{S}_{11})^{-\frac{1}{2}}\sqrt{n}\bm{Q}_{1n}(\bm{\theta}_0, \v{0})\right] + o_p(1).
\end{align*}
Let $\bm{Z} := (-\bm{S}_{11})^{-1/2} \sqrt{n} \bm{Q}_{1n}(\bm{\theta}_0, \v{0}) \dto \mathcal{N}(\v{0}, \bm{I})$ and $\bm{M} := (-\bm{S}_{11})^{-1/2} \bm{S}_{12} \bm{S}_{21}^{*} \bm{S}_{21} (-\bm{S}_{11})^{-1/2}$. Since $\bm{M}$ is symmetric and idempotent, with trace equal to $p$, the empirical likelihood ratio statistic $-2\log R_G(\bm{\theta}_0)/m = \bm{Z}^{'} \bm{M}\bm{Z} + o_p(1)$ converges to $\chi_p^2$.
\end{proof}

\subsection{Proof of Theorem 3}

\begin{proof}
Define the sample moment functions
\[
\bm{G}_{N} (\boldsymbol{\theta}_x) = \frac{1}{N} \sum_{i=1}^{N} \v{g}(\boldsymbol{X}_i, \boldsymbol{\theta}_x), \quad
\bm{G}_{M} (\boldsymbol{\theta}_y) = \frac{1}{M} \sum_{j=1}^{M} \v{g}(\boldsymbol{Y}_j, \boldsymbol{\theta}_y).
\]
At the true parameter values $\boldsymbol{\theta}_{x_0}$ and $\boldsymbol{\theta}_{y_0}$, by the law of large numbers (with Conditions 6 and 7 ensuring uniform integrability), we have
\[
\bm{G}_{N}(\boldsymbol{\theta}_{x_0}) = O_p(N^{-\frac{1}{2}}), \quad
\bm{G}_{M}(\boldsymbol{\theta}_{y_0}) = O_p(M^{-\frac{1}{2}}).
\]
By applying a first-order Taylor expansion of (\ref{6}) around $\boldsymbol{\lambda} = \v{0}$, and using the positive definiteness of the covariance matrices in Condition 5, we can get
\[
\boldsymbol{\lambda}^* = O_p(N^{-\frac{1}{2}}).
\]
Note that
\[
-2\log R_G(\bm\pi_0) = 2m\left\{\sum_{i=1}^{n_1}\log\left(1-\tau_1 (\boldsymbol{\lambda}^*)^{'}\bar{\v{g}}(\boldsymbol{X}_i, \boldsymbol{\theta}_x^*)\right) + \sum_{i=j}^{n_2}\log\left(1-\tau_2 (\boldsymbol{\lambda}^*)^{'}\bar{\v{g}}(\boldsymbol{Y}_j, \boldsymbol{\theta}_y^*)\right) \right\},
\]
expanding the log terms in $W_E(\boldsymbol{\pi}_0)$ using Taylor series yields
\begin{align*}
\log \left( 1 - \tau_1 (\boldsymbol{\lambda}^*)^{'} \bar{\v{g}}(\boldsymbol{X}_i, \boldsymbol{\theta}_x^*) \right) &=
-\tau_1 (\boldsymbol{\lambda}^*)^{'} \bar{\v{g}}(\boldsymbol{X}_i, \boldsymbol{\theta}_x^*) - \frac{1}{2} \left( \tau_1 (\boldsymbol{\lambda}^*)^{'} \bar{\v{g}}(\boldsymbol{X}_i, \boldsymbol{\theta}_x^*) \right)^2 + r_{1i} \\
\log \left( 1 + \tau_2 (\boldsymbol{\lambda}^*)^{'} \bar{\v{g}}(\boldsymbol{Y}_j, \boldsymbol{\theta}_y^*) \right) &=
\tau_2 (\boldsymbol{\lambda}^*)^{'} \bar{\v{g}}(\boldsymbol{Y}_j, \boldsymbol{\theta}_y^*) - \frac{1}{2} \left( \tau_2 (\boldsymbol{\lambda}^*)^{'} \bar{\v{g}}({Y}_j, \boldsymbol{\theta}_y^*) \right)^2 + r_{2j},
\end{align*}
where $|r_{1i}| \leq C_1 \left| \tau_1 (\boldsymbol{\lambda}^*)^{'} \bar{\v{g}}(\boldsymbol{X}_i, \boldsymbol{\theta}_x^*) \right|^3$ and
$|r_{2j}| \leq C_2 \left| \tau_2 (\boldsymbol{\lambda}^*)^{'} \bar{\v{g}}(\boldsymbol{Y}_j, \boldsymbol{\theta}_y^*) \right|^3$. Therefore
\begin{align*}
-2\log R_G(\boldsymbol{\pi}_0) &= 2m \left\{
\sum_{i=1}^{n_1} \left\{ -\tau_1 (\boldsymbol{\lambda}^*)^{'} \bar{\v{g}}(\boldsymbol{X}_i, \boldsymbol{\theta}_x^*) - \frac{\tau_1^2}{2} \left( (\boldsymbol{\lambda}^*)^{'} \bar{\v{g}}(\boldsymbol{X}_i, \boldsymbol{\theta}_x^*) \right)^2 \right\}
\right. \\
&\quad \left. + \sum_{j=1}^{n_2} \left\{ \tau_2 (\boldsymbol{\lambda}^*)^{'} \bar{\v{g}}(\boldsymbol{Y}_j, \boldsymbol{\theta}_y^*) - \frac{\tau_2^2}{2} \left( (\boldsymbol{\lambda}^*)^{'} \bar{\v{g}}(\boldsymbol{Y}_j, \boldsymbol{\theta}_y^*) \right)^2 \right\} + r
\right\},
\end{align*}
where $ r = \sum_{i=1}^{n_1} r_{1i} + \sum_{j=1}^{n_2} r_{2j}$. From (\ref{6}), we have
\[
\sum_{j=1}^{n_2} \frac{\bar{\v{g}}(\boldsymbol{Y}_j, \boldsymbol{\theta}_y^*)}{1 + \tau_2 (\boldsymbol{\lambda}^*)^{'} \bar{\v{g}}(\boldsymbol{Y}_j, \boldsymbol{\theta}_y^*)}
= \sum_{j=1}^{n_2} \bar{\v{g}}(\boldsymbol{Y}_j, \boldsymbol{\theta}_y^*) - \tau_2 \sum_{j=1}^{n_2} \bar{\v{g}}(\boldsymbol{Y}_j, \boldsymbol{\theta}_y^*) \bar{\v{g}}(\boldsymbol{Y}_j, \boldsymbol{\theta}_y^*)^{'} \boldsymbol{\lambda}^* + o_p(1) = \v{0},
\]
\[
\Rightarrow\sum_{j=1}^{n_2} \bar{\v{g}}(\boldsymbol{Y}_j, \boldsymbol{\theta}_y^*) = \tau_2 n_2 \left\{ \frac{1}{n_2} \sum_{j=1}^{n_2} \bar{\v{g}}(\boldsymbol{Y}_j, \boldsymbol{\theta}_y^*) \bar{\v{g}}(\boldsymbol{Y}_j, \boldsymbol{\theta}_y^*)^{'} \right\} \boldsymbol{\lambda}^* = n_2 \widetilde{\boldsymbol{V}}_2 \boldsymbol{\lambda}^*,
\]
where $\widetilde{\boldsymbol{V}}_2 = \tau_2\operatorname{Var}(\bar{\v{g}}(\boldsymbol{Y}, \boldsymbol{\theta}_{y_0}))$. Similarly, we have
\[
\sum_{i=1}^{n_1} \bar{\v{g}}(\boldsymbol{X}_i, \boldsymbol{\theta}_x^*) = -n_1 \widetilde{\boldsymbol{V}}_1 \bm{\lambda}^*,
\]
where $\widetilde{\boldsymbol{V}}_1 = \tau_1 \operatorname{Var}(\bar{\v{g}}(\bm{X}, \bm{\theta}_{x_0}))$. Let $\widetilde{\boldsymbol{V}} = \widetilde{\boldsymbol{V}}_1 + \widetilde{\boldsymbol{V}}_2$. The linear term becomes
\begin{align*}
&-2m \sum_{i=1}^{n_1} \tau_1 (\boldsymbol{\lambda}^*)^{'} \bar{\v{g}}(\boldsymbol{X}_i, \boldsymbol{\theta}_x^*) + 2m \sum_{j=1}^{n_2} \tau_2 (\boldsymbol{\lambda}^*)^{'} \bar{\v{g}}(\boldsymbol{Y}_j, \boldsymbol{\theta}_y^*)\\
&= 2N (\boldsymbol{\lambda}^*)^{'} \widetilde{\boldsymbol{V}} \boldsymbol{\lambda}^*.
\end{align*}
Combining the expressions, we can obtain
\begin{align*}
-2\log R_G(\boldsymbol{\pi}_0) = &2N (\boldsymbol{\lambda}^*)^{'} \widetilde{\boldsymbol{V}} \boldsymbol{\lambda}^* - m (\boldsymbol{\lambda}^*)^{'}\\
&\cdot \left\{
\tau_2^2 \sum_{j=1}^{n_2} \bar{\v{g}}(\boldsymbol{Y}_j, \boldsymbol{\theta}_y^*) \bar{\v{g}}(\boldsymbol{Y}_j, \boldsymbol{\theta}_y^*)^{'} +
\tau_1^2 \sum_{i=1}^{n_1} \bar{\v{g}}(\boldsymbol{X}_i, \boldsymbol{\theta}_x^*) \bar{\v{g}}(\boldsymbol{X}_i, \boldsymbol{\theta}_x^*)^{'}
\right\} \boldsymbol{\lambda}^* + r.
\end{align*}
By Condition 5, we have:
\[
\frac{1}{n_1} \sum_{i=1}^{n_1} \bar{\v{g}}(\boldsymbol{X}_i, \boldsymbol{\theta}_{x_0}) \bar{\v{g}}(\boldsymbol{X}_i, \boldsymbol{\theta}_{x_0})^{'} \overset{P}{\longrightarrow}
\frac{1}{m} \operatorname{Var}(\v{g}(\boldsymbol{X}, \boldsymbol{\theta}_{x_0})),
\]
\[
\frac{1}{n_2} \sum_{j=1}^{n_2} \bar{\v{g}}(\boldsymbol{Y}_j, \boldsymbol{\theta}_{y_0}) \bar{\v{g}}(\boldsymbol{Y}_j, \boldsymbol{\theta}_{y_0})^{'} \overset{P}{\longrightarrow}
\frac{1}{m} \operatorname{Var}(\v{g}(\boldsymbol{Y}, \boldsymbol{\theta}_{y_0})).
\]
\begin{align*}
\Rightarrow \quad&m \tau_1^2 \sum_{i=1}^{n_1} \bar{\v{g}}(\boldsymbol{X}_i, \boldsymbol{\theta}_x^*) \bar{\v{g}}(\boldsymbol{X}_i, \boldsymbol{\theta}_x^*)^{'} = N \widetilde{\boldsymbol{V}}_1 + o_p(1),\\
&m \tau_2^2 \sum_{j=1}^{n_2} \bar{\v{g}}(\boldsymbol{Y}_j, \boldsymbol{\theta}_y^*) \bar{\v{g}}(\boldsymbol{Y}_j, \boldsymbol{\theta}_y^*)^{'} = N \widetilde{\boldsymbol{V}}_2 + o_p(1).
\end{align*}
Hence
\begin{align*}
-2\log R_G(\boldsymbol{\pi}_0) &= 2N (\boldsymbol{\lambda}^*)^{'} \widetilde{\boldsymbol{V}} \boldsymbol{\lambda}^* - (\boldsymbol{\lambda}^*)^{'} (N \widetilde{\boldsymbol{V}}_1 + N \widetilde{\boldsymbol{V}}_2) \boldsymbol{\lambda}^* + o_p(1) \\
&= N (\boldsymbol{\lambda}^*)^{'} \widetilde{\boldsymbol{V}} \boldsymbol{\lambda}^* + o_p(1).
\end{align*}
Define
\[
\bar{\bm{G}}_1 := \frac{1}{n_1} \sum_{i=1}^{n_1} \bar{\v{g}}(\boldsymbol{X}_i, \boldsymbol{\theta}_{x_0}), \quad
\bar{\bm{G}}_2 := \frac{1}{n_2} \sum_{j=1}^{n_2} \bar{\v{g}}(\boldsymbol{Y}_j, \boldsymbol{\theta}_{y_0}).
\]
Perform a Taylor expansion on the first equation in (\ref{6})
\[
\frac{1}{n_1} \sum_{i=1}^{n_1} \bar{\v{g}}(\boldsymbol{X}_i, \boldsymbol{\theta}_x^*) + \left\{
\tau_1 \frac{1}{n_1} \sum_{i=1}^{n_1} \bar{\v{g}}(\boldsymbol{X}_i, \boldsymbol{\theta}_x^*) \bar{\v{g}}(\boldsymbol{X}_i, \boldsymbol{\theta}_x^*)^{'} - \widetilde{\boldsymbol{V}}_1
\right\} \boldsymbol{\lambda}^* + \widetilde{\boldsymbol{V}}_1 \boldsymbol{\lambda}^* + O_p(\frac{1}{n_1}) = \v{0}.
\]
Then we can get
\begin{align*}
\boldsymbol{\lambda}^* &= -\widetilde{\boldsymbol{V}}_1^{-1} \left( \frac{1}{n_1} \sum_{i=1}^{n_1} \bar{\v{g}}(\boldsymbol{X}_i, \boldsymbol{\theta}_x^*) \right) + O_p(\frac{1}{n_1})\\
&= -\widetilde{\boldsymbol{V}}_1^{-1} \bar{\bm{G}}_1 + O_p(\frac{1}{n_1}).
\end{align*}
Similarly,
\[
\boldsymbol{\lambda}^* = \widetilde{\boldsymbol{V}}_2^{-1} \bar{\bm{G}}_2 + O_p(\frac{1}{n_2}).
\]
\begin{align*}
\Rightarrow (\widetilde{\boldsymbol{V}}_1 + \widetilde{\boldsymbol{V}}_2) \boldsymbol{\lambda}^*
  &= (\bar{\bm{G}}_2 - \bar{\bm{G}}_1) + o_p(1) \\
  &= \frac{m}{N} \left(
      \tau_2 \sum_{j=1}^{n_2} \bar{\v{g}}(\boldsymbol{Y}_j, \boldsymbol{\theta}_{y_0}) -
      \tau_1 \sum_{i=1}^{n_1} \bar{\v{g}}(\boldsymbol{X}_i, \boldsymbol{\theta}_{x_0})
    \right) + o_p(1).
\end{align*}
Define $\boldsymbol{V}_1 := \tau_1 \operatorname{Var}(\v{g}(\boldsymbol{X}, \boldsymbol{\theta}_{x_0}))$, $\boldsymbol{V}_2 := \tau_2 \operatorname{Var}(\v{g}(\boldsymbol{Y}, \boldsymbol{\theta}_{y_0}))$. Then we have $\boldsymbol{V}_1 = m \widetilde{\boldsymbol{V}}_1$, $\boldsymbol{V}_2 = m\widetilde{\boldsymbol{V}}_2$. Thus, $\boldsymbol{V} = \boldsymbol{V}_1 + \boldsymbol{V}_2 = m \widetilde{\boldsymbol{V}}$. Therefore
\begin{align*}
(\widetilde{\boldsymbol{V}}_1 + \widetilde{\boldsymbol{V}}_2) \boldsymbol{\lambda}^*& = \frac{m}{N} \left(
      \tau_2 \sum_{j=1}^{n_2}\frac{1}{m}\sum_{t=1}^{m} \v{g}(\boldsymbol{Y}_{jt}, \boldsymbol{\theta}_{y_0}) -
      \tau_1 \sum_{i=1}^{n_1} \frac{1}{m}\sum_{k=1}^{m}\v{g}(\boldsymbol{X}_{ik}, \boldsymbol{\theta}_{x_0})
    \right) + o_p(1) \\
& = \frac{1}{N} \left(
      \tau_2 \sum_{j=1}^{N_2} \v{g}(\boldsymbol{Y}_j, \boldsymbol{\theta}_{y_0}) -
      \tau_1 \sum_{i=1}^{N_1} \v{g}(\boldsymbol{X}_i, \boldsymbol{\theta}_{x_0})
    \right) + o_p(1).
\end{align*}
Define
\[
\boldsymbol{D}_N := \frac{1}{\sqrt{N}} \left(
      \tau_2 \sum_{j=1}^{N_2} \v{g}(\boldsymbol{Y}_j, \boldsymbol{\theta}_{y_0}) -
      \tau_1 \sum_{i=1}^{N_1} \v{g}(\boldsymbol{X}_i, \boldsymbol{\theta}_{x_0})
    \right),
\]
by the Central Limit Theorem (with Condition 8 ensuring the Lindeberg condition), we have
\[
\boldsymbol{D}_N \xrightarrow{d} \mathcal{N}(\v{0}, \boldsymbol{\Sigma}),
\]
where $\boldsymbol{\Sigma} = \lim_{N \to \infty} \operatorname{Var}(\boldsymbol{D}_N) = \boldsymbol{V}_1 + \boldsymbol{V}_2 = \boldsymbol{V}$. Then we can get
\begin{align*}
\boldsymbol{\lambda}^* &= \widetilde{\boldsymbol{V}}^{-1} \frac{\boldsymbol{D}_N}{\sqrt{N}} + o_p(1) \\
&= m \boldsymbol{V}^{-1} \frac{\boldsymbol{D}_N}{\sqrt{N}} + o_p(1).
\end{align*}
Therefore
\begin{align*}
-2\log R_G(\boldsymbol{\pi}_0) &= N (\boldsymbol{\lambda}^*)^{'} \widetilde{\boldsymbol{V}} \boldsymbol{\lambda}^* + o_p(1)\\
&= N \left( m \boldsymbol{V}^{-1} \frac{\boldsymbol{D}_N}{\sqrt{N}} \right)^{'} \left( \frac{\boldsymbol{V}}{m} \right) \left( m \boldsymbol{V}^{-1} \frac{\boldsymbol{D}_N}{\sqrt{N}} \right) + o_p(1)\\
&= m \boldsymbol{D}_N^{'} \boldsymbol{V}^{-1} \boldsymbol{D}_N + o_p(1).
\end{align*}
Since \( \boldsymbol{V}^{-1/2} \boldsymbol{D}_N \xrightarrow{d} \mathcal{N}(\v{0}, \boldsymbol{I}_p) \), it follows that:
\[
\boldsymbol{D}_N^{'} \boldsymbol{V}^{-1} \boldsymbol{D}_N = \| \boldsymbol{V}^{-1/2} \boldsymbol{D}_N \|^2 \xrightarrow{d} \chi_p^2.
\]
By Condition 4 and the fact that \( \boldsymbol{\lambda}^* = O_p(N^{-1/2}) \), we can get
\[
\sum_{i=1}^{n_1} |r_{1i}| + \sum_{j=1}^{n_2} |r_{2j}| = O_p(N^{-1/2}).
\]
From Conditions 6 and 7 together with the M-estimation theory, we have
\[
\boldsymbol{\theta}_y^* - \boldsymbol{\theta}_{y_0} = O_p(N^{-1/2}), \quad \boldsymbol{\theta}_x^* - \boldsymbol{\theta}_{x_0} = O_p(N^{-1/2}).
\]
Hence \( r = o_p(1) \). Therefore
\[
\frac{-2\log R_G(\boldsymbol{\pi}_0) }{m}\xrightarrow{d} \chi_p^2.
\]
\end{proof}

\end{document}